\documentclass[11pt, a4paper]{article}
\usepackage{jheparxiv}
\usepackage[utf8]{inputenc}
\usepackage{amsmath}
\usepackage{amsfonts}
\usepackage{amssymb}
\usepackage{latexsym}
\usepackage{mathrsfs}
\usepackage{braket}		
\usepackage{setspace}
\usepackage{graphicx}
\usepackage{color}
\usepackage{xcolor}
\usepackage{slashed}
\usepackage{twistor}
\usepackage[all]{xy}
\usepackage{tikz-cd}
\usepackage{mathtools}
\usepackage{tikz}

\usepackage{graphicx}
\newcommand{\intprod}{\mathbin{\raisebox{\depth}{\scalebox{1}[-1]{$\lnot$}}}}

\newcommand{\sa}{\mathsf{a}}

\newcommand{\rg}{\mathrm{g}}

\renewcommand{\d}{\mathrm{d}}

\newcommand{\qed}{\hfill \ensuremath{\Box}}

\newcommand{\eps}{\epsilon}

\newcommand{\Dbar}{\bar{D}}

\newcommand{\msf}[1]{\mathsf{#1}}

\newcommand{\bA}{\mathbf{A}}
\newcommand{\bD}{\mathbf{D}}

\title{Higher-spin Yang-Mills, amplitudes and self-duality}

\author[a]{Tim Adamo}
\author[b]{\& Tung Tran}

\affiliation[a]{School of Mathematics and Maxwell Institute for Mathematical Sciences \\
        University of Edinburgh, EH9 3FD, United Kingdom}
        
\affiliation[b]{Service de Physique de l’Univers, Champs et Gravitation \\
Universit\'e de Mons, 20 place du Parc, 7000 Mons, Belgium}

\emailAdd{t.adamo@ed.ac.uk}
\emailAdd{vuongtung.tran@umons.ac.be}

\abstract{The existence of interacting higher-spin theories is tightly constrained by many no-go theorems. In this paper, we construct a chiral, higher-spin generalization of Yang-Mills theory in flat space which avoids these no-go theorems and has non-trivial tree-level scattering amplitudes with some higher-spin external legs. The fields and action are complex, so the theory is non-unitary and parity-violating, yet we find surprisingly compact formulae for all-multiplicity tree-level scattering amplitudes in the maximal helicity violating (MHV) sector, where the two negative helicity particles have identical but arbitrary spin. This is possible because the theory admits a perturbative expansion around its self-dual sector. Using twistor theory, we prove the classical integrability of this self-dual sector and show that it can be described on spacetime by an infinite tower of interacting massless scalar fields. We also give a twistor construction of the full theory, and use it to derive the formula for the MHV amplitude.}

\begin{document}

\maketitle
  
\section{Introduction}  

The study of higher-spin theories is motivated by both practical and theoretical questions. Massive higher-spin particles play a phenomenological role in describing composite states (such as those occurring in nuclear resonances of QCD) and are a crucial part of the spectrum of string theories. In the massless case, constructing higher-spin gravitational theories is important  for understanding the landscape of possible theories of quantum gravity, and has important implications ranging from holography to conformal bootstrap (see \cite{Sorokin:2004ie,Bekaert:2004qos,Rahman:2015pzl,Giombi:2016ejx,Bekaert:2022poo,Ponomarev:2022vjb} for reviews). But on a purely theoretical level, one can view higher-spin theories as a playground to explore what is -- and is not -- possible in the general frameworks of classical and quantum field theory.

Over the last 50 years, it has become clear that this is a theoretical playground with many rules. The possible array of higher-spin theories is tightly constrained by many no-go theorems, both for asymptotically flat spacetimes (e.g., \cite{Weinberg:1964ew,Coleman:1967ad,Weinberg:1980kq}) and asymptotically (A)dS spacetimes (e.g., \cite{Maldacena:2011jn,Maldacena:2012sf,Sleight:2017pcz}) -- see~\cite{Bekaert:2010hw,Rahman:2015pzl,Didenko:2014dwa,Giombi:2016ejx} for overviews. In attempting to evade these no-go theorems, there is no free lunch: providing an explicit construction of a higher-spin theory (even at the classical level) usually requires abandoning at least one feature that we usually think of as desirable for a physically interesting theory, such as locality, unitarity, or manifest covariance. To date, most known examples of local, massless higher-spin theories are either (quasi-)topological~\cite{Blencowe:1988gj,Bergshoeff:1989ns,Pope:1989vj,Fradkin:1989xt,Metsaev:1991mt,Metsaev:1991nb,Campoleoni:2010zq,Henneaux:2010xg,Gaberdiel:2010pz,Gaberdiel:2012uj,Gaberdiel:2014cha,Grigoriev:2019xmp,Grigoriev:2020lzu,Ponomarev:2016lrm,Tsulaia:2022csz} or higher-spin extensions of conformal gravity with higher-derivative equations of motion~\cite{Tseytlin:2002gz,Segal:2002gd,Bekaert:2010ky}. Moreover, those higher-spin theories which are defined in flat space turn out to have trivial S-matrices due to the severe constraints imposed on the interactions by the infinite-dimensional higher-spin symmetry (cf., \cite{Joung:2015eny,Beccaria:2016syk,Ponomarev:2016lrm})\footnote{It should be noted that some higher-spin theories with trivial S-matrices do have non-trivial boundary correlation functions when considered in an AdS background (in the sense that they are not equal to the correlation functions of a free CFT). For instance, chiral higher-spin gravity in AdS$_4$ is conjectured to be holographically dual to (a closed subset of) Chern-Simons matter theories \cite{Skvortsov:2018uru,Sharapov:2022awp}}. Depending on the audience, the triviality of higher-spin scattering can either be an intriguing feature of -- or a compelling reason to ignore -- higher-spin theories. In any case, it is an interesting question to ask: Is there a higher-spin theory in flat space with non-trivial scattering amplitudes, and if so what properties must it possess in order to avoid the many familiar constraints and no-go theorems?

\medskip

In this paper, we study a chiral, higher-spin version of Yang-Mills theory in four-dimensional flat spacetime which has non-trivial tree-level scattering amplitudes. This theory has been partially constructed before in the literature~\cite{Ponomarev:2017nrr,Krasnov:2021nsq,Tran:2021ukl}, and our work builds on these previous investigations. For brevity, we refer to this theory as \emph{higher-spin Yang-Mills} (HS-YM): it has many non-standard properties, which allow it to evade the net of no-go results on higher-spin theories with non-trivial S-matrices. In particular, the spin-$s$ gauge potentials of the theory live in certain ``un-balanced'' spin-$s$ representations of the Lorentz group, which we refer to as \emph{chiral representations} for every spin $s>1$, making the resulting gauge potentials intrinsically chiral. There are two on-shell degrees of freedom (i.e., positive and negative helicity) at each spin in this theory, but only one gauge-invariant field strength, from which the Lagrangian is constructed. The built-in chiral representations mean that the fields, Lagrangian and action functional of the theory are complex-valued in real, Lorentzian Minkowski spacetime. 

Having a complex action means that the theory is non-unitary, and the chiral representations break parity invariance. On the one hand, this means that HS-YM fails to have basic properties usually required of physical theories. But on the other hand, this ensures that HS-YM falls outside the assumptions of practically every no-go theorem constraining higher-spin interactions and scattering amplitudes. The chirality of the theory and higher-spin symmetries mean that in exchange processes, only spin-1 positive helicity particles can contribute (this is related to gauge invariance of the scattering amplitudes), although negative helicity particles can have arbitrary spin. Furthermore, its lack of unitarity and parity are fairly mild: self-dual Yang-Mills, self-dual gravity and conformal fishnet theory are also non-unitary with complex Lagrangians, but nevertheless encode a rich array of physically relevant information (cf., \cite{Bern:1993qk,Mahlon:1993si,Bardeen:1995gk,Bern:1996ja,Bern:1998sv,Krasnov:2016emc,Costello:2021bah,Costello:2022wso,Bu:2022dis,Bittleston:2022nfr} and \cite{Gurdogan:2015csr,Chicherin:2022nqq}). In any case, at tree-level one is always free to consider the theory defined in complexified or alternative spacetime signatures (such as Euclidean or $(2,2)$-signature), where the notion of complex fields and Lagrangians is less problematic.

In practical terms, we show that HS-YM has non-trivial scattering amplitudes by explicitly calculating the tree-level four-point amplitude using the Feynman rules of the spacetime action. When there are two positive helicity and two negative helicity external states, we find a non-vanishing amplitude, with the spins of the negative helicity particles identical but otherwise arbitrary.   

\medskip

In~\cite{Krasnov:2021nsq} a \emph{self-dual} subsector of HS-YM was defined, and it is straightforward to show that the full theory admits a perturbative (i.e., small coupling) expansion around this subsector. We give a description of self-dual HS-YM in terms of twistor theory~\cite{Penrose:1967wn,Ward:1977ta}, showing that it is classically integrable. A spacetime manifestation of this is the fact that self-dual HS-YM can be described by an infinite tower of massless, adjoint-valued scalar fields with cubic interactions; this is a higher-spin generalization of the Chalmers-Siegel description of self-dual Yang-Mills theory~\cite{Chalmers:1996rq}, which has already been written down as a contraction of chiral higher-spin gravity~\cite{Ponomarev:2017nrr}.  

These facts have several important consequences. Firstly, it means that HS-YM can be perturbatively expanded around a classically integrable subsector where we have vanishing tree amplitudes. Theories with such a structure can often be described in terms of \emph{twistor actions}\footnote{Examples include ordinary Yang-Mills theory~\cite{Mason:2005zm,Boels:2006ir}, conformal gravity~\cite{Mason:2005zm,Adamo:2013tja}, general relativity~\cite{Sharma:2021pkl}, 3-dimensional Yang-Mills-Higgs theory~\cite{Adamo:2017xaf} and conformal higher-spin theory~\cite{Haehnel:2016mlb,Adamo:2016ple}, as well as deformations of $\cN=4$ super-Yang-Mills and conformal fishnet theory~\cite{Adamo:2019lor}. There are also twistor constructions for the interactions of chiral higher-spin gravity~\cite{Tran:2022tft} and higher-spin generalizations of the IKKT matrix mode~\cite{Ishibashi:1996xs,Sperling:2017dts,Sperling:2018xrm,Steinacker:2022jjv}.}, classical reformulations of the theory on twistor space which have enhanced gauge invariance, that is a powerful tool for computing scattering amplitudes, and HS-YM is no exception. Furthermore, recent results on covariantizing chiral higher-spin theories~\cite{Metsaev:1991mt,Metsaev:1991nb,Ponomarev:2016lrm,Ponomarev:2016cwi,Ponomarev:2017nrr} using twistor-inspired methods and free differential algebras~\cite{Krasnov:2021nsq,Sharapov:2022faa,Sharapov:2022awp,Sharapov:2022wpz,Sharapov:2022nps} hint that twistor theory an ideal framework for constructing local higher-spin theories like HS-YM.

For scattering amplitudes in a helicity grading, the maximal helicity violating (MHV) configuration, with two negative helicity and arbitrarily many positive helicity external states, represents the first non-trivial perturbation away from self-duality and can be computed to all multiplicity directly from the twistor action of HS-YM. Remarkably, this leads to a compact formula for the $n$-point, color-ordered tree-level MHV amplitude of HS-YM written in spinor-helicity variables:
\be\label{introMHV}
\cA_{n}^{\mathrm{MHV}}=\frac{\rg^{n-2}}{2}\,\delta^{4}\!\left(\sum_{a=1}^{n}k_a\right)\,\frac{\la i\,j\ra^{2s+2}}{\la1\,2\ra\,\la2\,3\ra\cdots\la n\,1\ra}\,,
\ee 
where $\rg$ is the (dimensionless) coupling constant of the theory; $k^{\mu}_a$ is the on-shell 4-momentum of the $a^{\mathrm{th}}$ external particle; and the negative helicity particles $i,j$ have integer spin $s\geq1$ while all others have positive helicity and spin-1. While this formula can be guessed from $n=3,4$ explicit calculations and checked using BCFW recursion~\cite{Britto:2005fq}, the twistor action provides a first-principles derivation of the MHV amplitude.

\medskip

Besides the explicit construction of HS-YM on spacetime and calculation of tree-level scattering amplitudes, our main results can be summarized as follows:
\begin{itemize}
    \item[] \emph{Theorem~\ref{HS-WC}:} There is a one-to-one correspondence between solutions of the self-dual HS-YM equations and certain holomorphic vector bundles on twistor space, implying the classical integrability of the self-dual sector.
    
    \item[] \emph{Theorem~\ref{SDThm}:} The self-dual sector of HS-YM is described on spacetime by the action 
    \be
\frac{1}{2}\,\sum_{s=1}^{\infty}\,\int\tr\left(\d\phi^{(s)}\wedge *\d\phi^{(s)}\right) 
+\frac{1}{3}\sum_{s=1}^{\infty}\,\int\mu_{a,a}\wedge\tr\left(\phi^{(s)}\,\sum_{r+t=s+1}\d\phi^{(r)}\wedge\d\phi^{(t)}\right)\,,
\ee
where $\{\phi^{(s)}\}_{s=1,\ldots,\infty}$ are scalar functions valued in the adjoint representation of the gauge group and $\mu_{a,a}:=a_{\alpha}\,a_{\beta}\,\d x^{\alpha\dot\alpha}\wedge\d x^{\beta}{}_{\dot\alpha}$ for $a_{\alpha}$ some constant spinor.

  \item[] \emph{Theorem~\ref{TAThm}:} The classical action of full HS-YM theory is equivalent to an action functional on twistor space which has a local piece corresponding to the self-dual sector and a non-local piece encoding non-self-dual interactions.  
\end{itemize}

\medskip

The paper is structured as follows. Section \ref{sec:2} provides a definition of the spacetime theory for HS-YM and analyses higher-spin propagating degrees of freedom in the chiral representation. Section \ref{sec:3} computes the 3- and 4-point scattering amplitudes of HS-YM using Feynman rules before presenting the $n$-point formula \eqref{introMHV} for tree-level MHV scattering. Section \ref{sec:4} investigates the properties of self-dual HS-YM using twistor theory, establishing classical integrability of the self-dual sector and providing descriptions of it both on twistor space and spacetime. In Section \ref{sec:5}, we give a twistor action description of full HS-YM and use it to derive our formula for the tree-level MHV amplitudes. Section \ref{sec:6} concludes, and Appendix \ref{app:A} provides a check on the MHV formula using BCFW recursion.

\medskip

\paragraph{Notation:} Throughout, we denote SL$(2,\C)$ spinor indices of negative chirality by $\alpha,\beta,\ldots$ $=0,1$ and SL$(2,\C)$ spinor indices of positive chirality by $\dot{\alpha},\dot{\beta},\ldots=\dot{0},\dot{1}$. Spinor indices are raised and lowered using the two-dimensional Levi-Civita symbols:
\be\label{shcon}
b^{\alpha}=\epsilon^{\alpha\beta}\,b_{\beta}\,, \qquad b_{\alpha}=b^{\gamma}\,\epsilon_{\gamma\beta}\,, \qquad \epsilon^{\alpha\beta}\,\epsilon_{\alpha\gamma}=\delta^{\alpha}_{\gamma}\,,
\ee
and likewise for dotted indices. We often make use of the spinor helicity notation for SL$(2,\C)$-invariant contractions of spinors:
\be\label{shcon2}
\la a\,b\ra:=a^{\alpha}\,b_{\alpha}\,, \qquad [c\,d]:=c^{\dot\alpha}\,d_{\dot\alpha}\,,
\ee
where these inner products are skew-symmetric. Totally symmetric combinations of spinor indices will be denoted by $(\alpha_1\cdots\alpha_k)\equiv\alpha(k)$, $(\dot\alpha_1\cdots\dot\alpha_k)\equiv\dot\alpha(k)$, where symmetrization is always assumed to come with a prefactor of $\frac{1}{k!}$.

\medskip

\paragraph{Note added:} While this paper was being prepared, we became aware of~\cite{Herfray:2022prf}, which gives an interesting alternative construction of the self-dual sector of HS-YM in terms of non-projective twistor space.


\section{The space-time theory}\label{sec:2}

In four dimensions, a spin-$s$ gauge field is usually thought of as a totally symmetric rank-$s$ symmetric tensor~\cite{Fronsdal:1978rb}; exploiting the local isomorphism between the Lorentz group and SL$(2,\C)$, this is equivalent to representing the spin-$s$ gauge field by an object with $s$ un-dotted/negative chirality SL$(2,\C)$ spinor indices and $s$ dotted/positive chirality spinor indices. However, there are also `un-balanced' spin-$s$ representations of the gauge field, which have $2s$ total but unequal numbers of negative/positive chirality spinor indices. The price to be paid by working with such un-balanced representations is that they are not Lorentzian-real, as complex conjugation interchanges the spinor representations in Lorentzian signature, but in complexified spacetime or in Euclidean or split signature they are perfectly well-defined.

\begin{figure}[h!]
    \centering
    \includegraphics[scale=0.37]{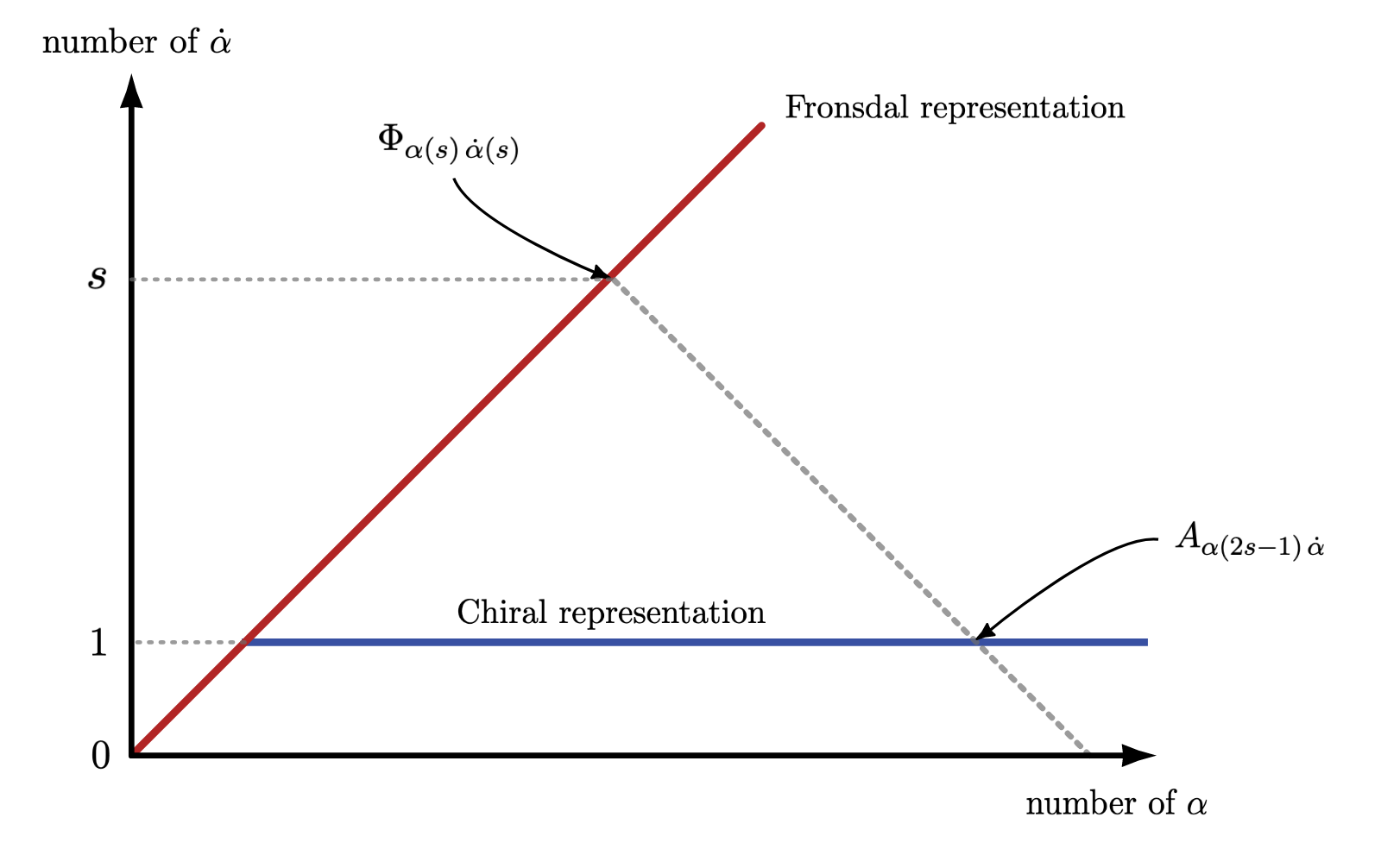}
    \caption{The standard Fronsdal representation (red) versus the chiral representation (blue) of a spin-$s$ gauge potential.}
\end{figure}
Following~\cite{Krasnov:2021nsq}, we will be interested in the un-balanced representation which has $2s-1$ un-dotted spinor indices and a single dotted spinor index for each integer spin $s\geq1$. We refer to this as the \emph{chiral representation}, although it has also been called by other names (`maximally un-balanced' or `twistor' representations) in the literature. We construct a theory whose field content is a higher-spin generalization of the Yang-Mills gauge potential in the chiral representation:   
\be\label{FCont}
\left\{A_{\alpha\dot\alpha}(x),\, A_{(\alpha\beta\gamma)\dot\alpha}(x),\,\ldots\right\}=\bigcup_{s=1}^{\infty}\left\{A_{\alpha(2s-1)\dot\alpha}\right\}\,,
\ee
where each of these potentials is valued in the adjoint of some simple Lie algebra $\mathfrak{g}$. As the notation suggests, for each value of $s$ the associated potential is totally symmetric in its negative chirality spinor indices, and since it is only these un-dotted spinor indices which proliferate when $s>1$, the theory is intrinsically chiral.

In this section, we review the basic classical structure of this higher-spin Yang-Mills (HS-YM) theory on space-time, including its field content, gauge symmetries and degrees of freedom. We note that aspects of this theory have appeared before in the literature: an action for the self-dual sector was given in~\cite{Krasnov:2021nsq}, and some features of the full theory were identified in~\cite{Tran:2021ukl}.


\subsection{Fields \& action}

From now on, we assume that we are working either on complexified Minkowski space $\M$, or\footnote{It is easy to extend all of our results to $\R^{2,2}$ as well, but we do not focus on this case here.} Euclidean $\R^4$. A standard method for compactly encoding higher-spin fields is to introduce an auxiliary commuting SL$(2,\C)$ spinor\footnote{For Euclidean reality conditions, these become SU$(2)$ spinors.} $y^\alpha$ and consider ``master'' gauge potentials which are polynomials in these auxiliary parameters/variables. We will adopt a slightly different (but completely equivalent) strategy which treats the master gauge potential not as a polynomial in $y^{\alpha}$ but as a homogeneous section of a bundle over $\M$. 

To do this, we introduce an auxiliary commuting SL$(2,\C)$ spinor $\lambda^{\alpha}$ which is considered only up to projective rescalings: that is, we identify $\lambda^{\alpha}\sim r\,\lambda^{\alpha}$ for any $r\in\C^*$. This is equivalent to viewing the projective equivalence class $[\lambda^\alpha]$ as homogeneous coordinates on the Riemann sphere $\P^1$. In this projective framework, a generic polynomial in $\lambda^\alpha$ is not well-defined (as it has no fixed weight under the projective scaling). Therefore, we will require the master gauge field to be a section of the \emph{infinite jet bundle} of the holomorphic tangent bundle of the Riemann sphere, $J^{\infty}(T\P^1)$, which is homogeneous of degree zero under projective rescalings. We will abuse notation slightly by abbreviating this bundle to $J^{\infty}_{\P^1}$, and also using this to denote the space of sections of the bundle.

In this work, we define HS-YM to be a theory of a non-abelian gauge connection
\be\label{HSconn1}
\bD_{\alpha\dot\alpha}=\partial_{\alpha\dot\alpha}+\bA_{\alpha\dot\alpha}(x|\lambda)\,,
\ee
where the connection 1-form takes values in $\Omega^{1}(\M)\otimes\mathfrak{g}\otimes J^{\infty}_{\P^1}$. Explicitly, this means that $\bA_{\alpha\dot\alpha}$ has an expansion of the form 
\be\label{ConJet0}
\bA_{\alpha\dot\alpha}(x|\lambda)=\sum_{s=1}^{\infty}\cA_{\beta(2s-2)|\alpha\dot\alpha}(x)\,\lambda^{\beta(2s-2)}\,\partial_0^{s-1}\,,
\ee
where the spacetime fields $\{\cA_{\beta(2s-2)|\alpha\dot\alpha}\}$ are valued in some Lie algebra $\mathfrak{g}$ and $\partial_0$ is the generator of the holomorphic tangent bundle of $\P^1$ (i.e., the section which trivializes the holomorphic tangent bundle). Note that $\lambda^{\beta(2s-2)}$ is a convenient notation for $\lambda^{(\beta_1}...\lambda^{\beta_{2s-2})}$. Since $T\P^1\cong\cO(2)$ as holomorphic line bundles, it follows that $\partial_0$ has weight $-2$ in $\lambda^\alpha$, and thus each term in \eqref{ConJet0} is homogeneous of degree zero in $\lambda^\alpha$, as required\footnote{More explicitly, the representation of $\partial_0$ provided by treating $\P^1$ as the Riemann sphere with positive-definite metric is:
\be\label{P1vec}
\partial_0:=\frac{\hat{\lambda}_{\alpha}}{\la \lambda\,\hat{\lambda}\ra}\,\frac{\partial}{\partial \lambda_{\alpha}}\,,
\ee
where $\hat{\lambda}^\alpha:=(\bar{\lambda}^{1},\,-\bar{\lambda}^0)$ is the antipodal point on $\P^1$.}. 

In order to define the action of the connection $\bD_{\alpha\dot\alpha}$ on objects valued in $\mathfrak{g}\otimes J^{\infty}_{\P^1}$, we first define a (somewhat trivial) Lie bracket for sections of $\mathfrak{g}\otimes J^{k}_{\P^1}$ for any $k\in\N$:
\be\label{HSbracket}
[\![f\partial_0^{a},\,g\partial_0^{b}]\!]:=[f,\,g]\,\partial_0^{a+b}\,,
\ee
where $f,g$ are Lie-algebra valued and $[\cdot\,,\cdot]$ is the usual Lie bracket on $\mathfrak{g}$. It is easy to check that \eqref{HSbracket} is itself a Lie bracket, and the connection acts on any adjoint-valued section $\Phi$ of $J^{\infty}_{\P^1}$ as
\be\label{ConAction}
\bD_{\alpha\dot\alpha}\Phi:=\partial_{\alpha\dot\alpha}\Phi+[\![\bA_{\alpha\dot\alpha},\,\Phi]\!]\,.
\ee
This enables us to define a field strength associated to the higher-spin gauge connection:
\be\label{FS1}
\mathbf{F}_{\alpha\beta}(x|\lambda):=\frac{\epsilon^{\dot\alpha\dot\beta}}{2}\,[\![\bD_{\alpha\dot\alpha},\,\bD_{\beta\dot\beta}]\!]\in\Omega^2_{-}(\M)\otimes\mathfrak{g}\otimes J^{\infty}_{\P^1}\,,
\ee
where $\Omega^{2}_{-}(\M)$ are the anti-self-dual (ASD) 2-forms on $\M$. In particular, we only consider the ASD part of the curvature associated with the partial connection.

Under non-abelian gauge transformations
\be\label{GT1}
\bA_{\alpha\dot\alpha}\to \mathbf{g}\,\bA_{\alpha\dot\alpha}\,\mathbf{g}^{-1}-\partial_{\alpha\dot\alpha}\mathbf{g}\,\mathbf{g}^{-1}\,,
\ee
where $\mathbf{g}$ is valued in $\mathfrak{g}\otimes J^{\infty}_{\P^1}$, the field strength transforms covariantly, $\mathbf{F}_{\alpha\beta}\to\mathbf{g}\,\mathbf{F}_{\alpha\beta}\,\mathbf{g}^{-1}$, as expected. It is easy to see that the non-ASD parts of the curvature of $\mathbf{D}$ do not transform covariantly with respect to these gauge transformations as a result of the underlying chirality of the construction (i.e., growing higher-spin degrees of freedom in un-dotted chiral representations, but not dotted ones). Thus, we see that the construction is doubly chiral: using the chiral representation (un-dotted) means that one only obtains sensible field strength components of corresponding chirality (ASD).

As it stands, this setup contains too many higher-spin gauge potentials. To see this, simply expand the coefficients $\cA_{\beta(2s-2)|\alpha\dot\alpha}(x)$ into un-dotted SL$(2,\C)$ irreducibles:
\be\label{cAdecompose}
\cA_{\beta(2s-2)|\alpha\dot\alpha}(x)= A_{(\beta(2s-2)\alpha)\dot\alpha}(x)+\epsilon_{\alpha(\beta_1}\,\msf{A}_{\beta(2s-3))\dot\alpha}(x)\,, \quad \forall s>1\,.
\ee
That is, for $s\geq2$, the master gauge field contains not one, but two non-abelian spin-$s$ gauge potentials: $A_{\alpha(2s-1)\dot\alpha}$ (the desired content) and $\msf{A}_{\beta(2(s+1)-3)\dot\alpha}$. It is easy to see that the superfluous field content decouples from any theory constructed from the $\mathbf{F}_{\alpha\beta}$, however. Indeed, $\msf{A}_{\beta(2s-3)\dot\alpha}$ drops out of $\mathbf{F}_{\alpha\beta}$, and $\mathbf{F}_{\alpha\beta}$ is left invariant by the local shift transformations
\be\label{Ashift}
\cA_{\beta(2s-2)|\alpha\dot\alpha}(x)\to\cA_{\beta(2s-2)|\alpha\dot\alpha}(x)+\epsilon_{\alpha(\beta_1}\,\vartheta_{\beta(2s-3))\dot\alpha}(x)\,,
\ee
which can be used to remove all of the $\msf{A}_{\beta(2s-3)\dot\alpha}$ components of the partial connection.

So without loss of generality, the master gauge potential can be taken to contain exactly the field content \eqref{FCont}:
\be\label{ConJet}
 \bA_{\alpha\dot\alpha}(x|\lambda):=\sum_{s=1}^{\infty} A_{\beta(2s-2)\alpha\dot\alpha}\,\lambda^{\beta(2s-2)}\,\partial_0^{s-1}\,,
\ee
as desired. This means that the component expansion of $\mathbf{F}_{\alpha\beta}$ is always totally symmetric in its un-dotted spinor indices, so that
\be\label{Fdecomp}
\mathbf{F}_{\alpha\beta}(x|\lambda)=\sum_{s=1}^{\infty}F_{(\alpha\beta\gamma(2s-2))}(x)\,\lambda^{\gamma(2s-2)}\,\partial_0^{s-1}\,,
\ee
with curvature components at each spin $s\geq1$ given by:
\be\label{ASDcurv}
F_{\alpha(2s)}:=\partial_{(\alpha_1}{}^{\dot\gamma}A_{\alpha(2s-1))\dot\gamma}+\sum_{r+t=s+1}\left[A_{(\alpha(2r-1)}{}^{\dot\gamma},\,A_{\alpha(2t-1))\dot\gamma}\right]\,.
\ee
Note that when $s=1$, this story truncates to the usual spinor description of (the ASD part of) a Yang-Mills gauge field. However, for $s>1$ the various higher-spin degrees of freedom mix with each other through the commutator terms: the gauge potential of spin $s>1$ will generate source terms in the field strengths at $s'>s$.

\medskip

Up to this point, the discussion has been purely kinematical, but we are now ready to define classical HS-YM with a spacetime action functional. For this, we require one additional structure, which is a M\"obius-invariant inner product on sections of $J^{\infty}_{\P^1}$ -- this is virtually identical to the inner product on the polynomial ring $\C[y^\alpha]$ introduced in~\cite{Krasnov:2021nsq}. Let
\be\label{innerprod0}
f=\sum_{a=0}^{\infty}f_{\alpha(2a)}\,\lambda^{\alpha(2a)}\,\partial_0^{a}\,, \qquad  g=\sum_{b=0}^{\infty}g_{\alpha(2b)}\,\lambda^{\alpha(2b)}\,\partial_0^{b}\,,
\ee
be any two section of $J^{\infty}_{\P^1}$. The required inner product is defined as:
\be\label{innerprod}
\left\langle \cdot\,|\,\cdot\right\rangle:J^{\infty}_{\P^1}\times J^{\infty}_{\P^1}\to\C\,, \qquad \left\langle f|g\right\rangle:=\sum_{a=1}^{\infty}\epsilon^{\alpha(2a)\beta(2a)}\,f_{\alpha(2a)}\,g_{\beta(2a)}\,.
\ee
Here, $\eps^{\alpha(2a)\beta(2a)}=\eps^{\alpha_{1}\beta_{1}}\cdots\eps^{\alpha_{2a}\beta_{2a}}$ with symmetrization over $\alpha$ and $\beta$ groups of indices, respectively. 

Armed with this, the HS-YM action is given by
\be\label{HS-YM}
S[\bA]=-\frac{1}{\rg^2}\int_\M \d^{4}x\,\tr\left\langle\mathbf{F}_{\alpha\beta}|\mathbf{F}^{\alpha\beta}\right\rangle= -\frac{1}{\rg^2}\,\sum_{s=1}^{\infty}\int_{\M}\d^{4}x\,\tr\left(F_{\alpha(2s)}\,F^{\alpha(2s)}\right)\,,
\ee
where $\rg$ is the dimensionless coupling constant and $\tr(\cdots)$ denotes the trace in $\mathfrak{g}$ (i.e., over the adjoint of the gauge group). Restricting to the $s=1$ sector returns a chiral action which is perturbatively equivalent to standard Yang-Mills theory, as they differ only by a topological term~\cite{Chalmers:1996rq}.

There is a nice property of HS-YM which is easily observed from this classical action, namely, that it admits a perturbative expansion around the \emph{self-dual sector}~\cite{Krasnov:2021nsq}
\be\label{SDHSYM}
F_{\alpha(2s)}=0\,, \quad \mbox{ for all } s=1,\ldots,\infty\,.
\ee
To see this, the action \eqref{HS-YM} can be re-written by introducing a set of higher-spin Lagrange multiplier fields
\be\label{Lmult}
\bB_{\alpha\beta}(x|\lambda):=\sum_{s=1}^{\infty}B_{(\gamma(2s-2)|\alpha\beta)}(x)\,\lambda^{\gamma(2s-2)}\,\partial_0^{s-1}\in\Omega^{2}_{-}(\M)\otimes\mathfrak{g}\otimes J^{\infty}_{\P^1}\,,
\ee
as
\be\label{ChSe1}
S[\bA,\bB]=\int_{\M}\d^{4}x\,\tr\left\langle\bB_{\alpha\beta}|\mathbf{F}^{\alpha\beta}\right\ra +\frac{\rg^2}{4}\,\int_{\M}\d^{4}x\,\tr\left\langle\bB_{\alpha\beta}|\bB^{\alpha\beta}\right\ra\,.
\ee
Note that we do not include any terms which are not totally symmetric in the expansion \eqref{Lmult} of $\bB_{\alpha\beta}$; this is because such terms decouple from the action when the gauge potential has the form \eqref{ConJet}. At the level of field components, the action \eqref{ChSe1} is simply
\be\label{ChSe2}
S[\bA,\bB]=\sum_{s=1}^{\infty}\int_{\M}\d^{4}x\,\tr\left(B_{\alpha(2s)}\,F^{\alpha(2s)}\right) +\frac{\rg^2}{4}\,\int_{\M}\d^{4}x\,\tr\left(B_{\alpha(2s)}\,B^{\alpha(2s)}\right)\,,
\ee
with field equations
\be\label{CSeq}
\mathbf{F}_{\alpha\beta}=-\frac{\rg^2}{2}\,\bB_{\alpha\beta}\,, \qquad \bD^{\alpha\dot\alpha}\bB_{\alpha\beta}=0\,.
\ee
Note that the Lagrange multipliers $\bB_{\alpha\beta}$ can be integrated out to return \eqref{HS-YM}. When $\rg^2\to0$ these equations reduce to the self-duality equations \eqref{SDHSYM}, along with a set of linear non-SD fields (given by $B_{\alpha(2s)}$) propagating on the SD HS-YM background. In Section~\ref{sec:4}, we will study the SD sector of HS-YM in some detail, showing that it is classically integrable and admits a twistor correspondence, as well as deriving action functionals (in twistor space and on spacetime) for the purely SD sector.


\subsection{Linear theory}

To get a better feel for the structure of HS-YM, it is instructive to look at the linear theory, which is described by the series of free spin-$s$ actions
\be\label{HSfree1}
S_{\mathrm{free}}[\bA]=\frac{1}{\rg^2}\,\sum_{s=1}^{\infty}\int_{\M}A^{\alpha(2s-1)\dot\alpha}\,\Box A_{\alpha(2s-1)\dot\alpha}\,,
\ee
after an integration-by-parts, where $\Box$ is the wave operator. At the linear level, the action is preserved by the gauge transformations
\be\label{FreeGT}
A_{\alpha(2s-1)\dot\alpha}\rightarrow A_{\alpha(2s-1)\dot\alpha}+\partial_{(\alpha_{1}|\dot\alpha|}\xi_{\alpha(2s-2))}\,,
\ee
and one can proceed to count the on-shell degrees of freedom by imposing a Lorenz gauge condition
\be\label{LorG}
\partial^{\alpha_1\dot\alpha}A_{\alpha_1\beta(2s-2)\dot\alpha}=0\,.
\ee
This removes $2s-1$ degrees of freedom from the $4s$ initially present in $A_{\alpha(2s-1)\dot\alpha}$, leaving $2s+1$. However, residual gauge transformations which obey $\Box\xi_{\alpha(2s-2)}=0$ leave the Lorenz gauge \eqref{LorG} intact, so this removes a further $2s-1$ degrees of freedom, leaving only \emph{two} on-shell degrees of freedom for HS-YM at each spin $s\geq1$. 

This means that rather than working with on-shell polarizations, we can label free HS-YM fields by their \emph{helicity}. However, the underlying chirality of HS-YM means that there is an asymmetry in the definition of positive and negative helicity. A \emph{positive helicity}, spin-$s$ HS-YM free field is a gauge potential $A^{(+)}_{\alpha(2s-1)\dot\alpha}$ whose linearized ASD curvature vanishes:
\be\label{ph1}
 \partial_{(\alpha_1}{}^{\dot\gamma}A^{(+)}_{\alpha(2s-1))\dot\gamma}=0\,.
\ee
On the other hand, a \emph{negative helicity}, spin-$s$ HS-YM free field is defined by a linearized ASD curvature $F^{(-)}_{\alpha(2s)}$ which obeys the negative helicity zero-rest-mass (z.r.m.) equation:
\be\label{nh1}
\partial^{\alpha\dot\alpha}F^{(-)}_{\alpha\beta(2s-1)}=0\,.
\ee
It should be noted that this sort of asymmetric definition is similar to what is encountered when characterizing helicity states in chiral background fields~\cite{Mason:2009afn,Adamo:2020yzi,Adamo:2022mev}. 

Momentum eigenstate representations for positive and negative helicity HS-YM fields will be useful when studying the scattering amplitudes of the theory. Let $k^{\alpha\dot\alpha}=\kappa^{\alpha}\tilde{\kappa}^{\dot\alpha}$ be an on-shell, massless (complex) 4-momentum. It is natural to follow the pattern for $s=1$ helicity states in the spinor-helicity formalism and define~\cite{Krasnov:2021nsq}:
\be\label{YMhel}
A^{(+)}_{\alpha(2s-1)\dot\alpha}=\frac{\zeta_{\alpha(2s-1)}\,\tilde{\kappa}_{\dot\alpha}}{\kappa^{\alpha_1}\cdots\kappa^{\alpha_{2s-1}}\,\zeta_{\alpha(2s-1)}}\,\e^{\im\,k\cdot x}\,, \qquad A^{(-)}_{\alpha(2s-1)\dot\alpha}=\frac{\kappa_{\alpha_1}\cdots\kappa_{\alpha_{2s-1}}\,\tilde{\zeta}_{\dot\alpha}}{[\tilde{\kappa}\,\tilde{\zeta}]}\,\e^{\im\,k\cdot x}\,,
\ee
where $\zeta_{\alpha(2s-1)},\tilde{\zeta}_{\dot\alpha}$ are constant spinors which obey $[\tilde{\zeta}\,\tilde{\kappa}]\neq0$ and $\zeta^{\beta\alpha(2s-2)} \kappa_{\beta}\neq0$. It is easy to show that these states obey \eqref{ph1} and \eqref{nh1}, respectively\footnote{Note that for the linear gauge potentials \eqref{YMhel} to have mass-dimension 1 (as required for the theory to have a single, dimensionless coupling constant, $\rg$) a mass scale must be present in the helicity polarizations for $s>1$~\cite{Krasnov:2021nsq}. We implicitly absorb this into the constant spinors $\zeta_{\alpha(2s-1)}$ and $\tilde{\zeta}_{\dot\alpha}$ throughout.}.

For the negative helicity state, it is obvious that the choice of $\tilde{\zeta}_{\dot\alpha}$ is pure gauge, as it drops out of the linearized ASD field strength:
\be\label{nh2}
F_{\alpha(2s)}[A^{(-)}]=\im\,\kappa_{\alpha_1}\cdots\kappa_{\alpha_{2s}}\,\e^{\im\,k\cdot x}\,.
\ee
For the positive helicity state, it is clear that $A^{(+)}$ is independent of the scale of $\zeta$, and along with the non-degeneracy condition ($\zeta^{\beta\alpha(2s-2)}\kappa_{\beta}\neq0$), this leaves exactly the residual gauge freedom contained in \eqref{FreeGT} after fixing Lorenz gauge (cf., \cite{Kaparulin:2012px}). In particular, this means that the choice of $\zeta_{\alpha(2s-1)}$ is \emph{not} pure gauge.

Indeed, it is easy to show that the difference between two $A^{(+)}$s with the same momentum but different choices of $\zeta_{\alpha(2s-1)}$ is not a gauge transformation \eqref{FreeGT}. Furthermore, the only gauge-invariant that can be formed from $A^{(+)}$ vanishes, by the positive helicity condition \eqref{ph1}. The only exception to these facts is when $s=1$, in which case the field is a positive helicity gluon and the choice of $\zeta_{\alpha}$ \emph{is} pure gauge. 

\medskip

This has important consequences for scattering amplitudes of the theory: in general, the requirement of gauge invariance means that only spin-1 positive helicity states can be involved, whereas negative helicity states are well-defined for arbitrary spin. Once again, this imbalance arises from the intrinsic chirality of the theory: in HS-YM, the only gauge-covariant field strength for $s>1$ is the ASD one, $F_{\alpha(2s)}$.

However, the fact that the action \eqref{HS-YM} is gauge-invariant, with two on-shell degrees of freedom for arbitrary spin, makes the $s=1$ constraint for positive helicity fields somewhat puzzling. Concretely, this is linked with the explicit choice of helicity basis \eqref{YMhel}. One could imagine that this is simply not the most general choice of helicity polarizations, and that there is a better choice which extends in a gauge-covariant way to all spins and helicities. Unfortunately, it is hard to see how \eqref{YMhel} could be altered or improved. The choice of the negative helicity polarization seems to be the only one which is consistent with little group scaling and matches the $s=1$ case. The normalization constraint $\epsilon^{(+)}_{\alpha(2s-1)\dot\alpha}\epsilon^{(-)\,\alpha(2s-1)\dot\alpha}=-1$, needed to recover the completeness relation for the polarization basis (see, e.g., the earlier work~\cite{Krasnov:2020bqr} where this unbalanced representation of polarization vectors was introduced), then essentially fixes the positive helicity polarization to be that given by \eqref{YMhel}. While not a \emph{proof} excluding some alternative helicity basis which allows for higher-spin positive helicity degrees of freedom in the theory, this line of reasoning does seem very constraining.

\medskip

Finally, to compute exchanges it will be necessary to have the propagator for HS-YM fields. With the Lorenz gauge condition \eqref{LorG}, the only propagator is between positive and negative helicity states:
\be\label{prop}
\begin{split}
\la A^{(+)}_{\alpha(2s-1)\dot\alpha}(k)\,A^{(-)\,\beta(2s'-1)\dot\beta}(k')\ra&=\delta^{4}(k+k')\,\delta_{s,s'}\,\delta_{s,1}\frac{\delta_{(\alpha_1}^{(\beta_1}\cdots\delta_{\alpha_{2s-1})}^{\beta_{2s-1})}\,\delta_{\dot\alpha}^{\dot\beta}}{k^2}\\
 &=\delta^{4}(k+k')\,\delta_{s,1}\,\delta_{s',1}\frac{\delta_{\alpha_1}^{\beta_1}\,\delta_{\dot\alpha}^{\dot\beta}}{k^2}\,,
\end{split}
\ee
where the trivial color structure (given by the Killing form on $\mathfrak{g}$) is suppressed. The constraint that the positive helicity particle has spin-1 (a consequence of gauge invariance) of course collapses the propagator to the usual gluon propagator.


\section{Scattering amplitudes}\label{sec:3}

Armed with the spacetime action of HS-YM \eqref{HS-YM} and a helicity basis of momentum eigenstates for the external fields, we can now proceed to investigate the structure of scattering amplitudes for this theory. Since the Lagrangian itself is not real-valued in Lorentzian signature, it makes sense for us to work with complex kinematics, leading to non-vanishing tree-level 3-point amplitudes. The vertex structure of the theory and gauge invariance constrains the exchanges to have only spin one at higher points, although the negative helicity external particles can have arbitrary spin. The complexity of the action combined with the fact that interactions are always at most single-derivative means that various no-go theorems prohibiting scattering amplitudes with higher-spin external legs can be evaded.


\subsection{3-point amplitudes}

As the external legs of any tree-level scattering amplitude in HS-YM are labeled by a helicity, these amplitudes can be denoted by $\cM_{n}(1_{s_1}^{h_1},\ldots,n_{s_n}^{h_n})$, where $h_{i}=\pm$ denotes the helicity (positive or negative) of the $i^{\mathrm{th}}$ external particle. This means that tree amplitudes can be helicity-graded by the number of, say, negative helicity external particles. At 3-points, this means that there are four possible helicity configurations: $(+,+,+)$, $(-,+,+)$, $(-,-,+)$ and $(-,-,-)$. For unitary theories with Lorentzian kinematics, it follows that all tree-level 3-point amplitudes vanish for the trivial reason that 
\be\label{3pkin}
\sum_{i=1}^{3}k_{i}=\sum_{i=1}^{3}\kappa_{i}^{\alpha}\,\bar{\kappa}_{i}^{\dot\alpha}=0 \quad \Rightarrow \quad \la i\,j\ra\,[i\,j]=0\,, \:\: \forall i,j\in\{1,2,3\}\,,
\ee
so all possible kinematic invariants vanish. Note that for \emph{complex} kinematics, where the momenta $k_{i}^{\alpha\dot\alpha}=\kappa_{i}^{\alpha}\tilde{\kappa}_{i}^{\dot\alpha}$ and $\tilde{\kappa}_{i}^{\dot\alpha}$ is not the complex conjugate of $\kappa_{i}^{\alpha}$, 3-particle momentum conservation only requires that one chirality of kinematical invariants vanish: namely all contractions of the form $\la i\,j\ra$, or all of the contractions of the form $[i\,j]$. As a consequence, this allows for potentially non-vanishing 3-point scattering amplitude configurations (cf., \cite{Benincasa:2007xk}). For instance, in ordinary Yang-Mills, one has non-vanishing $(-,+,+)$ (i.e., `$\overline{\mbox{MHV}}$') and $(-,-,+)$ (i.e., `MHV') 3-point amplitudes with complex kinematics. For Lorentzian-real theories this analytic continuation plays an important role by giving data with which to seed recursion relations and construct higher-multiplicity scattering amplitudes with real kinematics~\cite{Britto:2005fq,Benincasa:2011kn,Benincasa:2011pg}. However, in a complex theory like HS-YM such complex kinematics are natural from the outset. 

\medskip
With this in mind, the tree-level 3-point amplitudes of HS-YM are given by evaluating the cubic terms in the classical action \eqref{HS-YM} with on-shell external wavefunctions; this cubic interaction is given by
\be\label{cubic}
\tilde{\delta}(s_1-s_2-s_3+1)\,\rg\,\int\d^{4}x\,\tr\left(\partial_{(\alpha_1}{}^{\dot\gamma} A_{\alpha(2s_1-1))\dot\gamma}\,\left[A^{\alpha(2s_2-1)\dot\beta},\,A^{\alpha(2s_3-1)}{}_{\dot\beta}\right]\right)\,,
\ee
with 
\be\label{Kron}
\tilde{\delta}(x):=\left\{\begin{array}{ll}
                         0 & \mbox{ if } x\neq0 \\
                         1 & \mbox{ if } x=0
                         \end{array}\right.\,,
\ee
a Kronecker delta. The constraint on the spins is required for the integrand to be well-defined. Evaluating this cubic interaction with the momentum eigenstates \eqref{YMhel} -- and recalling that the constant spinors associated with negative helicity particles can be chosen arbitrarily -- it is easy to see that both $\cM_{3}(1^+,2^+,3^+)$ and $\cM_{3}(1^-,2^-,3^-)$ vanish for the same reasons as in pure Yang-Mills theory. 

\medskip

This leaves only the MHV and $\overline{\mbox{MHV}}$ configurations as non-vanishing 3-point amplitudes. Although gauge invariance dictates that in general only spin-1 positive helicity states are allowed, for now we keep the spins arbitrary. In the $\overline{\mbox{MHV}}$ case, evaluating the cubic vertex on the momentum eigenstates leads in the first instance to:
\be\label{MHV-bar0}
\cM_{3}(1_{s_1}^-,2_{s_2}^+,3_{s_3}^+)=\im\,\rg\,f^{\sa_1\sa_2\sa_3}\,\tilde{\delta}(s_1-s_2-s_3+1)\,\frac{[2\,3]\,\la\zeta_2\,1\ra^{2s_2-1}\,\la\zeta_3\,1\ra^{2s_3-1}}{\la\zeta_2\,2\ra^{2s_2-1}\,\la\zeta_3\,3\ra^{2s_3-1}}\,,
\ee
where $f^{\mathsf{abc}}$ are the structure constants of the gauge group, the overall momentum conserving delta function has been stripped off and (without loss of generality) we have decomposed the positive helicity reference spinors as $$\zeta_{2}^{\alpha(2s_2-1)}=\zeta_2^{\alpha_1}\cdots\zeta^{\alpha_{2s_2-1}}_{2},$$ etc. Now, on the support of (complex) momentum conservation, it follows that
\be\label{momcon1}
\la\zeta_2\,1\ra\,[1\,3]+\la\zeta_2\,2\ra\,[2\,3]=0\,, \qquad \la\zeta_3\,1\ra\,[1\,2]+\la\zeta_3\,3\ra\,[3\,2]=0\,,
\ee
which means that the $\overline{\mbox{MHV}}$ amplitude is equal to
\be\label{MHV-bar}
\cM_{3}(1_{s_1}^-,2_{s_2}^+,3_{s_3}^+)=\im\,\rg\,f^{\sa_1\sa_2\sa_3}\,\tilde{\delta}(s_1-s_2-s_3+1)\,\frac{[2\,3]^{2s_1+1}}{[1\,2]^{2s_3-1}\,[3\,1]^{2s_2-1}}\,,
\ee
matching the formula found in~\cite{Krasnov:2021nsq} for the self-dual sector of HS-YM. 

Observe that the highly-constraining 3-point kinematics mean that the result is manifestly gauge-invariant \emph{for all spins} satisfying the constraint. This is an accident, unique to 3-point amplitudes (as we will soon see). Imposing the constraints $s_2=s_3=1$ from the start, the remaining spin constraint in \eqref{MHV-bar} sets $s_1=1$ and the whole $\overline{\mbox{MHV}}$ 3-point amplitude collapses to that of pure Yang-Mills.  

\medskip

The 3-point MHV amplitude is evaluated along similar lines, leading to
\begin{multline}\label{MHV}
\cM_{3}(1_{s_1}^{-},2_{s_2}^{-},3_{s_3}^{+})=\frac{\im\,\rg}{2}\,f^{\sa_1\sa_2\sa_3}\,\frac{\la1\,2\ra^{2s_3}}{\la2\,3\ra^{2s_3-1}\,\la3\,1\ra^{2s_3-1}}\Big[\la1\,2\ra^{2s_2-1}\,\la3\,1\ra^{2s_3-2}\,\tilde{\delta}(s_1-s_2-s_3+1)  \\
 + \la1\,2\ra^{2s_1-1}\,\la3\,2\ra^{2s_3-2}\,\tilde{\delta}(s_2-s_1-s_3+1)\Big]\,,
\end{multline}
where the constraint $s_3=1$ has been temporarily ignored. Here, the two terms arise from the need to symmetrize over the location of the positive helicity particle in the cubic vertex \eqref{cubic}. Once again, the constant spinor used to define the positive helicity polarization drops out of the amplitude, leaving an ``accidentally'' gauge-invariant result for all external spins. A striking thing about this MHV amplitude is that it is \emph{not}, for generic spins, the helicity conjugate of its $\overline{\mbox{MHV}}$ counterpart \eqref{MHV-bar}. This is, of course, an unavoidable consequence of the chirality of the theory, which leads to a violation of parity invariance.

When $s_3=1$ is imposed (as it should have been from the start), \eqref{MHV} simplifies to
\begin{equation}
\begin{split}
 \cM_{3}(1_{s_1}^{-},2_{s_2}^{-},3_{1}^{+})&=\frac{\im\,\rg}{2}\,f^{\sa_1\sa_2\sa_3}\,\frac{\la1\,2\ra^{2}}{\la2\,3\ra\,\la3\,1\ra}\Big[\la1\,2\ra^{2s_2-1}\,\tilde{\delta}(s_1-s_2) + \la1\,2\ra^{2s_1-1}\,\tilde{\delta}(s_2-s_1)\Big] \\
  & = \im\,\rg\,f^{\sa_1\sa_2\sa_3}\,\tilde{\delta}(s_1-s_2)\,\frac{\la1\,2\ra^{2s_1+1}}{\la2\,3\ra\,\la3\,1\ra}\,,
\end{split}
\end{equation}
where both negative helicity external particles must have identical -- but otherwise arbitrary -- spin. When $s_1=s_2=s_3=1$ the formula reduces to the 3-point MHV amplitude of pure Yang-Mills, which is the parity conjugate of the $\overline{\mbox{MHV}}$ with all spin-one external fields. The reason for this is that, when restricted to spin-one gauge fields, the action \eqref{HS-YM} differs from the Yang-Mills action only by a topological term, so parity invariance holds perturbatively despite the chirality of the Lagrangian~\cite{Chalmers:1996rq}. The same cannot be said of the chiral action of full HS-YM theory, which is clearly not perturbatively equivalent to any parity-invariant theory.


\subsection{4-point amplitudes}

Now, let us turn to the computation of 4-point tree-level scattering amplitudes in HS-YM. The cubic interactions are extended off-shell and linked together with the propagator \eqref{prop}, with the appropriate spin constraints at each vertex in any given Feynman diagram. In addition, we have contributions from the quartic contact interaction 
\be\label{quartic}
\tilde{\delta}(s_1+s_2-s_3-s_4)\,\rg^2\,\int\d^{4}x\,\tr\left(\Big[A_{(\alpha(2s_1-1)}{}^{\dot\gamma},\,A_{\alpha(2s_2-1))\dot\gamma}\Big]\,\left[A^{\alpha(2s_3-1)\dot\delta},\,A^{\alpha(2s_4-1)}{}_{\dot\delta}\right]\right)\,,
\ee
with the spin constraint ensuring that the spinor contractions are well-defined. Unlike the 3-point amplitudes, at this stage gauge invariance requires all positive helicity particles to have spin-1.

Once again, we can proceed by helicity-grading the amplitudes, but the calculation is further simplified by restricting our attention to color-ordered partial amplitudes. In particular, it is easy to show that tree-level scattering amplitudes decompose as
\be\label{colororder}
\cM_{n}(1^{h_1}_{s_1},\ldots,n^{h_n}_{s_n})=\rg^{n-2}\,\delta^{4}\!\left(\sum_{i=1}^{n}k_i\right) \sum_{\sigma\in S_{n}\setminus\Z_n}\tr(\mathsf{T}^{\sa_{\sigma(1)}}\cdots\mathsf{T}^{\sa_{\sigma(n)}})\,\cA_{n}(\sigma(1^{h_1}_{s_1}),\ldots,\sigma(n_{s_n}^{h_n}))\,,
\ee
in terms of a sum over distinct (i.e., non-cyclically related) color-orderings; here $\mathsf{T}^{\sa}$ are generators of the gauge group and $h_i=\pm$ is the helicity of the $i^{\mathrm{th}}$ particle. The functions of the kinematic data $\cA_n$ are the color-ordered partial amplitudes -- knowing $\cA_n$ in any color-ordering thus determines the full amplitude $\cM_n$.

Given that we only have non-vanishing 3-point MHV and $\overline{\mbox{MHV}}$ amplitudes, simple factorization arguments immediately indicate that $\cA_{4}(1^+,2^+,3^+,4^+)=0$, since the exchanges involved in such an amplitude vanish while the 4-point contact contribution can be eliminated by making appropriate gauge choices for the constant reference spinors in the external states. However, the next helicity configuration, $\cA_{4}(1^-,2^+,3^+,4^+)$, does not \emph{a priori} vanish. In this color-ordering, the amplitude receives contributions from exchange diagrams in the $s$- and $t$-channels, as well as a contact term\footnote{Where it is useful, we denote the spin of the $i^{\mathrm{th}}$ external particle in a scattering amplituded with a $s_i$ subscript in $\cA_n$.}:
\begin{equation*}
   \cA_{4}(1_{s_1}^-,2_{1}^+,3_{1}^+,4_{1}^+)= \widehat{\cA}_{4}^s+\widehat{\cA}_{4}^{t}+\widehat{\cA}_{4}^{\mathrm{cont}}\,.
\end{equation*}
We first compute the $s$-channel exchange:
\be\label{schannel}
    \widehat{\cA}_4^s=(-1)^{3-s_1}\,\frac{[12]^{2-s_1}\,[34]\,f(\zeta_2,\zeta_3,\zeta_4)}{(k_1+k_2)^2}\,\tilde{\delta}(1-s_1)\,,
\ee
where the rational function $f$ depends on the auxiliary spinors of the positive helicity fields:
\be
    f(\zeta_2,\zeta_3,\zeta_4):=\frac{\langle 
    \zeta_2\,1\rangle}{\langle \zeta_2\,2\rangle}\,\frac{\langle 
    \zeta_3\,4\rangle}{\langle 
    \zeta_3\,3\rangle}\,\frac{\langle 
    \zeta_4\,3\rangle}{\langle 
    \zeta_4\,4\rangle}\,\frac{\langle\zeta_3\,1\rangle\langle \zeta_4\,2\rangle}{\langle\zeta_3\,3\rangle\langle \zeta_4\,4\rangle}\,\left(\frac{\langle 
    \zeta_4\,1\rangle}{\langle 
    \zeta_4\,2\rangle}\right)^{s_1}\,,
\ee
which is homogeneous of weight zero in the reference spinors, as required.

Now, as the external positive helicity states are spin-1, choice of the reference spinors is just residual gauge freedom and we can set $\zeta_{2}^{\alpha}=\zeta_{3}^{\alpha}=\zeta_{4}^{\alpha}=\kappa_{1}^{\alpha}$, from which it immediately follows that $f(\zeta_2,\zeta_3,\zeta_4)=0$, and thus the $s$-channel contribution vanishes $\widehat{\cA}^s_4=0$. A similar calculation shows that the $t$-channel contribution also vanishes: $\widehat{\cA}^t_4=0$. The only remaining contributions are from the contact interaction; in this color-ordering the contact contributions are of the form
\be\label{4ptcontact}
\tilde{\delta}(s_1-1)\,\frac{\la1\,\zeta_2\ra^{2s_1-1}\,\la \zeta_4\,\zeta_3\ra\,\la\zeta_2\,\zeta_3\ra^{2(1-s_1)}\,[\tilde{\zeta}_1\,2]\,[3\,4]}{[1\,\tilde{\zeta}_1]\,\la2\,\zeta_2\ra\,\la3\,\zeta_3\ra\,\la4\,\zeta_4\ra} -\,(2\leftrightarrow3)\,.
\ee
Clearly, this contribution is always proportional to contractions of the form $\la1\,\zeta_i\ra$ (for $i\neq1$), which are killed with the residual gauge fixing $\zeta_i=\kappa_1$. 

Thus, it follows that the amplitude in this helicity configuration vanishes:
\be\label{1minus}
\cA_{4}(1_{s_1}^-,2_{1}^+,3_{1}^+,4_{1}^+)=0\,,
\ee
regardless of the spins of the external fields. Since the only vertices contributing to this amplitude are the $\overline{\mbox{MHV}}$ 3-point ones, the computation of this amplitude is the same as in the purely self-dual theory, and the vanishing of the amplitude is in agreement with light-cone results for the self-dual sector~\cite{Skvortsov:2018jea,Skvortsov:2020wtf,Skvortsov:2020gpn}.

\medskip

Next, we come to the 4-point MHV helicity configuration, with two negative and two positive helicity external fields. Let us begin by computing $\cA_{4}(1_{s_1}^-,2_{s_2}^-,3_{1}^+,4_{1}^+)$. Once again, in this color-ordering the exchanges are in the $s$- and $t$-channels; partially fixing the residual gauge symmetry so that
\be\label{pgf}
\zeta_{3}^{\alpha}=\zeta_{4}^{\alpha}=\zeta^{\alpha}\,,
\ee
subject to $\la\zeta\,3\ra\neq0\neq\la\zeta\,4\ra$, the $s$-channel contribution is given by:
\be\label{MHVschan}
\widehat{\cA}_{4}^{s}=\frac{(-1)^{2-s_1-s_2}}{2}\,\frac{\la1\,2\ra^{2s_1-2}\,\la\zeta\,2\ra\,[\tilde{\zeta}_1|k_1+k_2|\zeta\ra\,[3\,4]}{[2\,1]\,[1\,\tilde{\zeta}_1]\,\la3\,\zeta\ra\,\la4\,\zeta\ra}\,\tilde{\delta}(s_1-s_2)\,+\:(1\leftrightarrow2)\,.
\ee
Here, the remaining spin constraint fixes the two negative helicity particles to have identical spin, $s_1=s_2$, but otherwise their spin is unconstrained. Similar expressions arise for the $t$-channel exchange and contact diagram, all with the same spin constraint.

\medskip

Upon further fixing the gauge redundancy by setting
\be\label{pgf2}
\zeta^{\alpha}=\kappa_{1}^{\alpha}\,, \qquad \tilde{\zeta}_{1}^{\dot\alpha}=\tilde{\kappa}_{4}^{\dot\alpha}\,,
\ee
and exploiting 4-momentum conservation, the $t$-channel and contact contributions are easily seen to vanish, while $s$-channel contribution collapses to give the full amplitude
\be\label{4pMHV1}
\cA_{4}(1_{s}^-,2_{s}^-,3_{1}^+,4_{1}^+)=\frac{\la1\,2\ra^{2s+1}}{\la2\,3\ra\,\la3\,4\ra\,\la4\,1\ra}
\ee
for the 4-point MHV amplitude in this color-ordering.

\medskip

The fact that \eqref{4pMHV1} is non-vanishing for generic higher spins $s>1$ raises the alarm: aren't we violating well-known no-go theorems constraining S-matrices with higher-spin external states? As alluded to above, the basic properties of HS-YM theory mean that no-go theorems (e.g., Weinberg's low energy theorem~\cite{Weinberg:1964ew}, Weinberg-Witten~\cite{Weinberg:1980kq}, Coleman-Mandula~\cite{Coleman:1967ad}, etc.) simply do not apply. In particular, the theory is purely massless, contains no scalars, is parity-violating, non-unitary and its interactions have at most one derivative. Furthermore, the exchanges themselves are spin-1, so in effect the MHV amplitude is corresponding to two negative helicity higher-spin fields interacting with a positive helicity pure gluon background. Various subgroups of these properties violate the assumptions of all no-go theorems constraining the tree-level S-matrix.

\medskip

For completeness, we provide the expression for the MHV amplitude of HS-YM in the color-ordering where the negative helicity particles are not consecutive. Following similar steps to above, one arrives at the formula
\be\label{4pMHVac*}
\cA_{4}(1_{s_1}^-,2_{1}^+,3_{s_3}^-,4_{1}^+)=\tilde{\delta}(s_1-s_3)\,\frac{\la1\,3\ra^{2s_1+2}}{\la1\,2\ra\la2\,3\ra\,\la3\,4\ra\,\la4\,1\ra}\,.
\ee
Once again, the spins of the external negative helicity particles are identical but otherwise arbitrary.


\subsection{$n$-point MHV amplitudes}

Based on the pattern observed at 4-points, it is tempting to conjecture an \emph{all-multiplicity} formula for the tree-level scattering amplitudes of HS-YM in the MHV helicity configuration (i.e., two negative helicity higher-spin particles and arbitrarily many positive helicity external gluons). The natural conjecture is:
\be\label{nMHV}
\cA_{n}(1^{+}_{1},\ldots,i^{-}_{s_i},\ldots,j^{-}_{s_j},\ldots,n^{+}_{1})=\tilde{\delta}(s_i-s_j)\,\frac{\la i\,j\ra^{2s_i+2}}{\la1\,2\ra\,\la2\,3\ra\cdots\la n\,1\ra}\,,
\ee
where particles $i,j$ have negative helicity. This formula passes several basic consistency checks: it reduces to the well-known Parke-Taylor formula for $n$-gluon MHV scattering~\cite{Parke:1986gb} when $s_1=\cdots=s_n=1$, carries the correct little group weight in each external particle and has only the usual collinear poles of ordinary Yang-Mills theory.

While directly computing this formula from the Feynman rules of HS-YM is clearly not tractable, there are other ways of confirming that it is correct. In Appendix~\ref{app:A}, we confirm \eqref{nMHV} using BCFW recursion~\cite{Britto:2005fq} after first showing that HS-YM can indeed be constructed via on-shell recursion. This is possible because of the inherent chirality of the theory, which allows it to evade no-go factorization arguments for higher-spin theories~\cite{Benincasa:2007xk,Benincasa:2011kn,Benincasa:2011pg,McGady:2013sga}. In Section~\ref{sec:5}, we derive \eqref{nMHV} directly from the HS-YM action using twistor theory.

\medskip

Before concluding this section, it is worth illustrating, in practical terms, \emph{why} the restriction to spin-1 for positive helicity external particles is necessary. One way to test this is to calculate 4-point amplitudes using the assumption that the reference spinors $\zeta_{\alpha(2s-1)}$ can be arbitrarily chosen for $s>1$; that is, by assuming that gauge invariance will be respected. Performing the 4-point MHV calculation and then extrapolating to higher-multiplicity leads to the formula:
\begin{multline}\label{nMHVarb}
\cA_{n}(1^{+}_{s_1},\ldots,i^{-}_{s_i},\ldots,j^{-}_{s_j},\ldots,n^{+}_{s_n})=\frac{\la i\,j\ra^{4}}{\la1\,2\ra\,\la2\,3\ra\cdots\la n\,1\ra} \\ \times\left[\tilde{\delta}\!\left(s_j-s_i-n+2+\sum_{a\neq i,j}s_a\right) \,\left(\frac{\la i\,j\ra^{s_j-n+1+\sum_{a\neq i,j}s_a}}{\prod_{b\neq i,j} \la j\,b\ra^{s_b-1}}\right)^2\right. \\
\left.+\tilde{\delta}\!\left(s_i-s_j-n+2+\sum_{a\neq i,j}s_a\right) \,\left(\frac{\la i\,j\ra^{s_i-n+1+\sum_{a\neq i,j}s_a}}{\prod_{b\neq i,j} \la i\,b\ra^{s_b-1}}\right)^2\right]\,.
\end{multline}
At first, this may seem like a reasonable formula: it carries the correct little group weights, obeys the symmetries imposed by the color-ordering, and collapses to \eqref{nMHV} when $s_a=1$ for all $a\neq i,j$. Furthermore, when the two negative helicity particles are gluons ($s_i=1=s_j$), the spin constraints become
\be\label{nMHVarbsc}
\sum_{a\neq i,j}s_a=n-2 \quad \Rightarrow \quad s_a=1, 
\ee
for all $a\neq i,j$, since each $s_a\geq1$.

However, the formula \eqref{nMHVarb} now has higher-order poles whenever the negative helicity momenta become collinear with \emph{any} of the positive helicity momenta, regardless of their position in the color-ordering. These are not physical for a colored, two-derivative local field theory, and the root of these spurious singularities can be traced back precisely to identifying the positive helicity reference spinors with some external momenta.

It would, of course, be interesting to explore whether \eqref{nMHVarb} can be understood as a valid scattering amplitude in some non-local context, but this is beyond the scope of the current paper.


\section{Self-dual sector and integrability}\label{sec:4}

We have already identified a \emph{self-dual} (SD) sector of HS-YM theory, corresponding to the condition
\be\label{SDsec}
F_{\alpha(2s)}=0\,, \qquad \mbox{for all } s\geq1\,.
\ee
For the $s=1$ truncation of the theory, these are the familiar self-duality equations of Yang-Mills theory, which are known to be classically integrable using a variety of different perspectives (cf., \cite{Yang:1977zf,Atiyah:1977pw,Atiyah:1978ri,Mason:1991rf}). It is natural to ask if the SD sector of HS-YM is likewise classically integrable.

The scattering amplitude calculations of the previous section hint that this should be true, as we found that $\cA_{4}(1^+,2^+,3^+,4^+)$ and $\cA_{4}(1^-,2^+,3^+,4^+)$ vanish for HS-YM. This is indicative of a self-dual sector which is consistent and classically integrable, respectively.

In this section, we answer the question of the classical integrability of HS-YM in the affirmative using twistor theory. In particular, we generalize Ward's theorem~\cite{Ward:1977ta} for SD Yang-Mills theory to the higher-spin setting, proving an equivalence between all SD HS-YM fields and certain integrable holomorphic structures in twistor space. We then use this construction and holomorphic Chern-Simons theory in twistor space to arrive at a spacetime description for SD HS-YM as a four-dimensional theory of an infinite tower of adjoint-valued scalars.


\subsection{Twistor theory}

Penrose's twistor theory gives a non-local description of spacetime physics in terms of complex projective geometry~\cite{Penrose:1967wn}, and has now found many different uses across theoretical and mathematical physics. Rather than provide an extensive review of this rich subject, we give a brief recap of the features required for the study of HS-YM; in-depth reviews can be found in~\cite{Penrose:1972ia,Penrose:1985bww,Penrose:1986ca,Ward:1990vs,Mason:1991rf,Dunajski:2010zz}, and we follow the notation of~\cite{Adamo:2017qyl}.

Let $\M$ be complexified Minkowski spacetime; the real spacetimes of various signatures -- Lorentzian $\R^{1,3}$, Euclidean $\R^4$ and Kleinian $\R^{2,2}$ -- sit inside this complexified spacetime as real slices. The (projective) twistor space $\PT$ of $\M$ is given by an open subset of three-dimensional complex projective space $\P^3$
\be\label{PT}
\PT=\left\{Z^{A}=(\mu^{\dot\alpha},\lambda_{\alpha})\in\P^3\,|\,\lambda_{\alpha}\neq0\right\}\,,
\ee
where 
\be\label{homcoord}
Z^{A}\sim r\,Z^{A}\,, \quad \forall r\in\C^{*}\,,
\ee
are homogeneous coordinates on $\P^3$ defined only up to this projective rescaling. We will denote the equivalence class of such homogeneous coordinates under projective rescaling as $[Z^A]$. Since $\lambda_{\alpha}\neq0$ on $\PT$, there is a natural fibration
\be\label{fibration}
\pi:\PT\to\P^1\,, \qquad [Z^{A}]\mapsto[\lambda_{\alpha}]\,,
\ee
with $\lambda_{\alpha}$ serving as homogeneous coordinates on the $\P^1$ base of the fibration. 

The correspondence between $\PT$ and $\M$ is given by the incidence relations
\be\label{increl}
\mu^{\dot\alpha}=x^{\alpha\dot\alpha}\,\lambda_{\alpha}\,,
\ee
which state that each point $x\in\M$ corresponds to a holomorphic, linearly embedded Riemann sphere $X\cong\P^1\subset\PT$. Conversely, any point $Z^{A}=(\mu^{\dot\alpha},\lambda_{\alpha})$ in twistor space corresponds to a totally null ASD 2-plane in $\M$, whose tangent vectors have the form $\lambda^{\alpha}v^{\dot\alpha}$ for fixed $\lambda^{\alpha}$ (given by the choice of $Z^A$) and arbitrary $v^{\dot\alpha}$. These totally null ASD 2-planes are called $\alpha$-planes in $\M$.   

There are many interesting results which follow from this basic non-local geometric correspondence between $\PT$ and $\M$. For our purposes, there are two classic results which will prove most important. The first of these is the \emph{Penrose transform}, which gives an equivalence between solutions of the massless free field (or zero-rest-mass) equations on $\M$ of any integer or half-integer spin, and cohomology classes on $\PT$~\cite{Penrose:1969ae,Eastwood:1981jy,Baston:1989vh}. More precisely, this takes the form of an isomorphism:
\be\label{PenTran}
\left\{\mbox{massless free fields on }\M \mbox{ of helicity } h\right\} \cong H^{0,1}(\PT,\cO(2h-2))\,,
\ee
where it is assumed that the set of massless free fields comes with some suitable regularity conditions and $H^{0,1}(\PT,\cO(2h-2))$ denotes the Dolbeault cohomology\footnote{Of course, any realization of the cohomology group $H^{1}(\PT,\cO(2h-2))$ will do, but we find it most useful to work with the Dolbeault representation, following~\cite{Woodhouse:1985id}.} group of $(0,1)$-forms on $\PT$ valued in $\cO(2h-2)$, the sheaf of holomorphic homogeneous functions of weight $2h-2$.

The second major result of twistor theory that is crucial for us is the \emph{Ward correspondence}, which gives a one-to-one correspondence between solutions of the SD Yang-Mills equations on $\M$ and certain holomorphic vector bundles over $\PT$~\cite{Ward:1977ta}. Our first result is to generalize this correspondence to the SD sector of HS-YM theory.


\subsection{Twistor construction of self-dual HS-YM}

As one might expect, there is a higher-spin version of the Ward correspondence~\cite{Ward:1977ta} for the SD sector of HS-YM:

\begin{thm}\label{HS-WC}
 There is a one-to-one correspondence between:
 \begin{itemize}
  \item self-dual HS-YM connections with gauge group GL$(N,\C)$, and

  \item holomorphic bundles $V=E\otimes J^{\infty}_{\P^1}\rightarrow\PT$, where $E$ is a rank $N$ bundle which is topologically trivial on restriction to any line in $\PT$ and $J^{\infty}_{\P^1}$ is identified with the infinite jet bundle of the bundle of horizontal vectors of the fibration $\pi:\PT\to\P^1$.
 \end{itemize}
\end{thm}

\proof First, suppose that we are given a self-dual, GL$(N,\C)$ HS-YM field on $\M$; this is characterized by the equations \eqref{SDsec} for each $s\geq1$. Let $\alpha_Z$ denote the $\alpha$-plane $\M$ corresponding to some point $Z\in\PT$; tangent vectors of $\alpha_Z$ have the form $\lambda^{\alpha}v^{\dot\alpha}$ for fixed $\lambda^{\alpha}$. Due to the chirality of HS-YM, only the ASD part of the field strength has a gauge invariant definition, and this is set to zero by SD equations. However, on restriction to $\alpha_Z$, it follows that any SD HS-YM connection obeys
\be\label{apl}
[\![\mathbf{D}_{\alpha\dot\alpha},\,\mathbf{D}_{\beta\dot\beta}]\!]\Big|_{\alpha_Z}=\lambda^{\alpha}\,\lambda^{\beta}\,v^{\dot\alpha}\,w^{\dot\beta}\,\frac{\epsilon_{\alpha\beta}}{2}\,[\![\mathbf{D}^{\gamma}{}_{\dot\alpha},\,\mathbf{D}_{\gamma\dot\beta}]\!]=0\,.
\ee
In other words, on restriction to an $\alpha$-plane, the SD HS-YM connection is totally flat -- there is no ambiguity in defining the SD part of the field strength because its restriction to the $\alpha$-plane vanishes. Thus, the set of covariantly constant sections valued in the fundamental representation of GL$(N,\C)$ is a set of constant functions.

Next, define the vector space
\be\label{Wardfibre}
V|_{Z}=\left\{\mathfrak{s}(x|\lambda) \mbox{ valued in } \C^N\otimes J^{\infty}_{\P^1}\,|\,\bD_{\alpha\dot\alpha}\mathfrak{s}=0\right\}\cong \C^N\otimes J^{\infty}_{\P^1}\,,
\ee
and making the identification between the auxiliary projective spinor $\lambda_{\alpha}$ on $\M$ and the coordinate on the base of the twistor fibration $\pi:\PT\to\P^1$. This provides a holomorphic construction of the fibres of a vector bundle $V=E\otimes J^{\infty}_{\P^1}\rightarrow\PT$ of appropriate rank, and by construction this bundle will be topologically trivial upon restriction to any twistor line $X$.

For the converse, given the vector bundle $V\rightarrow\PT$, the condition of holomorphicity is equivalent to the bundle being endowed with a partial connection
\be\label{pconn1}
\Dbar: \Omega^0(\PT, V)\rightarrow\Omega^{0,1}(\PT, V)\,, \qquad \Dbar^2=0\,.
\ee
Locally, this partial connection can be written in terms of a potential $\msf{a}\in\Omega^{0,1}(\PT,\mathrm{End}\,V)$, with
\be\label{pconn2}
\Dbar=\dbar+\msf{a}\,, \qquad \msf{a}=\sum_{s=1}^{\infty}a^{(s)}\,\partial_0^{s-1}\,, \qquad a^{(s)}\in\Omega^{0,1}(\PT,\mathrm{End}\,E\otimes\cO(2s-2))\,,
\ee
subject to
\be\label{tint}
F^{(0,2)}[\sa]=\dbar\msf{a}+[\![\msf{a},\,\msf{a}]\!]=0\,.
\ee
This is the condition that the partial connection on $V$ is holomorphic. Locally, the action on sections of $V$ is given by
\be\label{pconnact}
\Dbar\phi=\dbar\phi+[\![\msf{a},\,\phi]\!]\,,
\ee
for all $\phi\in\Omega^0(\PT,V)$.

The assumption of topological triviality on any twistor line $X$, combined with a ``sufficient smallness'' assumption on the data $\msf{a}$ implies that $V|_{X}$ can also be holomorphically trivialized (cf., \cite{Sparling:1990,Mason:2010yk}). This implies the existence of a holomorphic frame
\be\label{hframe}
H(x,\lambda,\bar{\lambda}):V|_{X}\rightarrow \C^{N}\otimes J^{\infty}_{\P^1}\,, \qquad \Dbar|_{X} H=0\,.
\ee
Any such holomorphic frame is only unique up to transformations of the form
\be\label{gtran}
H\rightarrow H\,\mathbf{g}(x,\lambda)\,, \qquad \mathbf{g}(x,\lambda)=\sum_{s=1}^{\infty}g_{\alpha(2s-2)}(x)\,\lambda^{\alpha(2s-2)}\,\partial_0^{s-1}\,,
\ee
where the coefficient functions $g_{\alpha(2s-2)}(x)$ are valued in the Lie algebra $\mathfrak{gl}_N$. 

In terms of the parametrization \eqref{pconn2}, the condition on the holomorphic frame reads
\be\label{hframe2}
\dbar|_{X}H+\msf{a}|_{X}\,H=0\,.
\ee
Since $\msf{a}$ is defined on twistor space, the incidence relations \eqref{increl} ensure that $\lambda^{\alpha}\partial_{\alpha\dot\alpha}\msf{a}|_X=0$. Furthermore, since $\lambda^{\alpha}\partial_{\alpha\dot\alpha}$ is a holomorphic vector field, it follows that
\be\label{hframe3}
\dbar|_{X}\left(H^{-1}\,\lambda^{\alpha}\partial_{\alpha\dot\alpha}\,H\right)=0\,.
\ee
Thus, $H^{-1}\,\lambda^{\alpha}\partial_{\alpha\dot\alpha}\,H$ is a holomorphic section of $\cO(1)\otimes\mathfrak{gl}_N\otimes J^{\infty}_{\P^1}$. By a straightforward extension of Liouville's theorem to weighted, bundle-valued functions, it follows that
\be\label{Sparlingt}
H^{-1}\,\lambda^{\alpha}\partial_{\alpha\dot\alpha}\,H=\lambda^{\alpha}\,\bA_{\alpha\dot\alpha}(x|\lambda)\,,
\ee
where 
\be\label{HSconnt}
\bA_{\alpha\dot\alpha}(x|\lambda):=\sum_{s=1}^{\infty} A_{\dot\alpha(\alpha\beta(2s-2))}(x)\,\lambda^{\beta(2s-2)}\,\partial_0^{s-1}\,.
\ee
Under a change of holomorphic frame \eqref{gtran}, it is easy to see that $\bD_{\alpha\dot\alpha}=\partial_{\alpha\dot\alpha}+\bA_{\alpha\dot\alpha}$ transforms as \eqref{GT1}\footnote{We have neglected to include the additional terms corresponding to the $\mathsf{A}_{\beta(2s-3)\dot\alpha}$ in \eqref{cAdecompose} in the decomposition, because these can be removed by gauge transformations, or, equivalently, rotations of the holomorphic frame on twistor space.}. Thus, we recover the field content of HS-YM in terms of the usual higher-spin gauge connection.

\medskip

The self-duality condition arises as a consequence of the integrability of the partial connection: $\Dbar^2=0$ imposes a constraint on the Lax pair $\lambda^{\alpha}\bD_{\alpha\dot\alpha}$, which is simply
\be\label{Laxint}
[\![\lambda^{\alpha}\bD_{\alpha\dot\alpha},\,\lambda^{\beta}\bD_{\beta\dot\beta}]\!]=0 \quad \Leftrightarrow \quad F_{\alpha(2s)}=0 \:\:\: \forall s\geq1\,.
\ee
Thus, we obtain the SD HS-YM equations from the holomorphic bundle $V\rightarrow\PT$, as desired. \qed

It is easy to adapt this theorem to other gauge groups, following the usual prescriptions (cf., \cite{Ward:1990vs}). For instance, to get gauge group SU$(N)$, one must supplement the conditions on $V$ with the requirement that it admit a positive real form and have trivial determinant line bundle. The theorem also descends to the real slices $\R^4$ and $\R^{2,2}$ (but not to real-valued fields on $\R^{1,3}$ due to the chirality of the theory and the SD sector). For Euclidean reality conditions, one requires the real form to be preserved under the anti-holomorphic involution which acts as the antipodal map on twistor lines, whereas for split signature the real form must descend to the $\RP^3$ real slice of twistor space.

This theorem also implies that:
\begin{corollary}
 The SD sector of HS-YM theory is classically integrable.
\end{corollary}

\proof This follows straightforwardly from the proof of Theorem~\ref{HS-WC}, which equates the SD equations \eqref{SDsec} with the integrability condition for an elliptic operator $\Dbar^2=0$ on twistor space. Equivalently, there is an integrable Lax pair associated with the SD sector, given by $\lambda^{\alpha}\bD_{\alpha\dot\alpha}$. \qed


\subsection{Action functional for the self-dual sector} 

The self-dual sector of pure Yang-Mills theory has many descriptions in terms of action functionals in four-dimensions which translate the classical integrability into a constrain on some auxiliary degrees of freedom. These formulations include the Chalmers-Siegel action (written in terms of an adjoint-valued scalar)~\cite{Chalmers:1996rq} and a four-dimensional Wess-Zumino-Witten (WZW) model~\cite{Donaldson:1985zz}. Using the Ward correspondence, it turns out that these and many other spacetime actions for SD Yang-Mills can be derived by performing dimensional reductions from holomorphic Chern-Simons theories on twistor space~\cite{Costello:2020a,Bittleston:2020hfv,Costello:2021bah}; these theories require certain choices of boundary conditions to be well-defined. Different gauge choices in twistor space induce different spacetime descriptions.

In light of Theorem~\ref{HS-WC} it is natural to ask if similar constructions hold for self-dual HS-YM. As self-duality is equated with integrability of the partial connection $\Dbar$ on a bundle $E\otimes J^{\infty}_{\P^1}\to\PT$, the natural starting point is an action functional on $\PT$ whose only equation of motion is \eqref{tint}: $F^{(0,2)}[\msf{a}]=0$. These are precisely the equations of motion of a \emph{holomorphic Chern-Simons} theory for the partial connection $\Dbar$~\cite{Witten:1992fb,Thomas:1997}. In general, these theories are only well-defined on Calabi-Yau manifolds, where there is a global section of the canonical bundle to wedge against the holomorphic Chern-Simons form; since $\PT$ is not Calabi-Yau, making sense of the theory requires choosing some boundary conditions. Here, we only make one such choice; there are many others which would be interesting to investigate further.

Let us restrict our attention to Euclidean reality conditions, for which $\PT\cong\R^4\times\P^1$ and the incidence relations can be inverted
\be\label{invirel}
x^{\alpha\dot\alpha}=\frac{\hat{\mu}^{\dot\alpha}\,\lambda^{\alpha}-\mu^{\dot\alpha}\,\hat{\lambda}^{\alpha}}{\la\lambda\,\hat{\lambda}\ra}\,,
\ee
where $\hat{\lambda}^{\alpha}=(\bar{\lambda}^1,\,-\bar{\lambda}^0)$ and $\hat{\mu}^{\dot\alpha}=(\bar{\mu}^{\dot1},\,-\bar{\mu}^{\dot0})$. Useful bases for the holomorphic and anti-holomorphic tangent and cotangent bundles of twistor space are provided with these reality conditions by~\cite{Woodhouse:1985id}:
\be\label{holbas}
\partial_0=\frac{\hat{\lambda}_{\alpha}}{\la \lambda\,\hat{\lambda}\ra}\,\frac{\partial}{\partial \lambda_{\alpha}}\,, \quad \partial_{\dot\alpha}=-\frac{\hat{\lambda}^{\alpha}\,\partial_{\alpha\dot\alpha}}{\la\lambda\,\hat{\lambda}\ra}\,, \quad e^{0}=\la\lambda\,\d\lambda\ra\,, \quad e^{\dot\alpha}=\lambda_{\alpha}\,\d x^{\alpha\dot\alpha}\,,
\ee
and
\be\label{aholbas}
\dbar_0=-\la\lambda\,\hat{\lambda}\ra\,\lambda_{\alpha}\frac{\partial}{\partial\hat{\lambda}_{\alpha}}\,, \quad \dbar_{\dot\alpha}=\lambda^{\alpha}\partial_{\alpha\dot\alpha}\,, \quad \bar{e}^0=\frac{\la\hat{\lambda}\,\d\hat{\lambda}\ra}{\la\lambda\,\hat{\lambda}\ra^2}\,, \quad \bar{e}^{\dot\alpha}=\frac{\hat{\lambda}_{\alpha}\,\d x^{\alpha\dot\alpha}}{\la\lambda\,\hat{\lambda}\ra}\,,
\ee
respectively. With this in mind, we define a holomorphic Chern-Simons form
\be\label{hCSform}
\begin{split}
\mathrm{hCS}[\msf{a}] :&=\tr\left(\msf{a}\wedge\dbar\msf{a}+\frac{2}{3}\,\msf{a}\wedge[\![\msf{a}\wedge\msf{a}]\!]\right) \\
 &=\sum_{s=1}^{\infty}\tr\left(a^{(s)}\wedge\dbar a^{(s)}+\frac{2}{3}\,a^{(s)}\wedge\sum_{r+t=s+1}a^{(r)}\wedge a^{(t)}\right)\partial_0^{2s-2}\,,
\end{split}
\ee
which takes values in $\Omega^{0,3}(\PT,(J^{\infty}_{\P^1})^2)$, where $(J^{\infty}_{\P^1})^2$ denotes the infinite jet bundle whose sections are composed of only \emph{even} powers of $\partial_0$.

To form a holomorphic Chern-Simons action, we must wedge this against a section $\Omega$ of $\Omega^{3,0}(\PT,(J^{\infty\,\vee}_{\P^1})^2)$, where $J^{\infty\,\vee}_{\P^1}$ is the dual of the infinite jet bundle, generated by $e^0$ in \eqref{holbas}. The pairing by inner product $\iota_{\partial_0}(e^0)\equiv \p_0 \intprod e^0=1$ then eliminates all the generators of the infinite jet bundle and its dual, so that $\Omega\wedge\mathrm{hCS}[\msf{a}]$ is a $(3,3)$-form which makes sense to integrate over $\PT$. 

However, since the canonical bundle of $\PT$ is $\cO(-4)$ as a line bundle, some poles must be introduced to render $\Omega$ weightless. There are many possible choices (cf., \cite{Costello:2020a,Bittleston:2020hfv}), but here we consider:
\be\label{poltf}
\Omega:=\frac{\D^{3}Z}{\la a\,\lambda\ra^{4}}\,\sum_{s=1}^{\infty}\left(\frac{e^0}{\la a\,\lambda\ra^2}\right)^{2s-2}\,,
\ee
for $\D^{3}Z:=\epsilon_{ABCD}Z^{A}\d Z^{B}\d Z^{C}\d Z^{D}$ the weight $+4$ top holomorphic form on $\PT$. In other words, $\Omega$ is defined by having poles (starting at fourth-order) at $A^A=(0,a_{\alpha})\in \PT$ on twistor space. With this choice, the holomorphic Chern-Simons action  
\be\label{hCS}
S[\msf{a}]=\frac{1}{2\pi\im}\int_{\PT}\Omega\wedge\mathrm{hCS}[\msf{a}]\,,
\ee
is well-defined.

\medskip

Na\"ively, the field equations of this action are precisely $F^{(0,2)}[\msf{a}]=0$, as desired. However, the poles appearing in $\Omega$ mean that in order to have a well-defined variational problem associated with this action, the twistor gauge potential $\msf{a}(Z)$ must have zeros of appropriate order in each term of its infinite jet bundle expansion. In particular, we must have that
\be\label{bconds}
\msf{a}=\sum_{s=1}^{\infty}a^{(s)}\p_0^{s-1}=\sum_{s=1}^{\infty}\la a\,\lambda\ra^{2s}\,\varphi^{(s)}(Z)\,\partial_0^{s-1}\,,
\ee
where 
\be\label{tscalars}
\varphi^{(s)}\in\Omega^{0,1}(\PT,\cO(-2)\otimes\mathfrak{g})\,, \qquad \forall\:s\geq1\,,
\ee
which can be thought of as a boundary condition on $\msf{a}$ at the point $A^A=(0,a_{\alpha})\in\PT$. Likewise, infinitesimal gauge transformations of the form $\sa\rightarrow \sa+\bar{\p}\xi+[\![\sa,\xi]\!]$ must obey
\begin{align}
    \xi=\sum_{s=1}^{\infty}\langle a\,\lambda\rangle^{2s}\,\psi^{(s)}(Z)\,\p_0^{s-1}\,,\qquad \psi^{(s)}\in \Omega^{0}(\PT,\cO(-2)\otimes \mathfrak{g})\,,
\end{align}
to preserve this boundary condition.

Note that we can use this gauge freedom, and the generic existence of a holomorphic trivialization of the bundle $V\to\PT$, to make the partial connection restricted to any holomorphic curve $X$ pure gauge:
\be\label{BSgauge}
\msf{a}|_{X}=\msf{a}_0\,\bar{e}^0=\hat{\sigma}^{-1}\dbar|_{X}\hat\sigma\,,
\ee
where $\hat{\sigma}:\PT\to G\otimes J^{\infty}_{\P^1}$, for gauge group $G$ and $\hat{\sigma}^{-1}$ is understood to be an inverse only with respect to the gauge group factor. Now, in the expansion \eqref{bconds}, we can impose that each $\varphi^{(s)}$ is harmonic upon restriction to any twistor line, which implies that~\cite{Woodhouse:1985id}
\be\label{harmonic}
\varphi^{(s)}|_{X}=\bar{e}^{0}\,\phi^{(s)}(x)\,,
\ee
for $\{\phi^{(s)}(x)\}$ an infinite tower of adjoint-valued functions on $\R^4$, one for each spin in the spectrum of HS-YM.

This allows us to solve for the gauge transformation $\hat{\sigma}$ explicitly, taking
\be\label{hatsig}
\hat{\sigma}=\exp\left[-\sum_{s=1}^{\infty}\frac{\la a\,\lambda\ra^{2s-1}\,\la a\,\hat{\lambda}\ra}{\la\lambda\,\hat{\lambda}\ra}\,\phi^{(s)}\,\partial_0^{s-1}\right]\,,
\ee
where the residual gauge freedom is fixed by the boundary condition $\hat{\sigma}(x,a)=\mathrm{id}_{G}$. The partial connection in this gauge can then be written as
\be\label{BSgauge2}
\Dbar=\dbar+\msf{a}=\hat{\sigma}^{-1}\left(\dbar+\msf{a}'_{\dot\alpha}\,\bar{e}^{\dot\alpha}\right)\hat{\sigma}\,,
\ee
where the components of the field equation $F^{(0,2)}[\msf{a}]=0$ along the $\P^1$ fibres of the bundle $\PT\to\R^{4}$ dictate that 
\be\label{aprime}
\msf{a}'_{\dot\alpha}=\lambda^{\alpha}\,\sum_{s=1}^{\infty}A_{\beta(2s-2)\alpha\dot{\alpha}}(x)\,\lambda^{\beta(2s-2)}\,\partial_0^{s-1}\,,
\ee
for the set of $\{A_{\alpha(2s-1)\dot\alpha}\}$ adjoint-valued HS-YM gauge potentials on $\R^4$. 

Now, from \eqref{BSgauge2} it follows that
\be\label{BSgauge3}
\msf{a}_{\dot\alpha}=\hat{\sigma}^{-1}\left(\dbar_{\dot\alpha}+\msf{a}'_{\dot\alpha}\right)\hat{\sigma}\,,
\ee
and a straightforward calculation using \eqref{hatsig} and \eqref{aholbas} leads to
\begin{multline}\label{BSgauge4}
\msf{a}_{\dot\alpha}=\hat{\sigma}^{-1}\left[\la\lambda\,a\ra\,\sum_{s=1}^{\infty}\left(\la a\,\lambda\ra^{2s-2}\,a^{\alpha}\partial_{\alpha\dot\alpha}\phi^{(s)}-\hat{a}^{\alpha}\,A_{\alpha\beta(2s-2)\dot\alpha}\,\lambda^{\beta(2s-2)}\right)\,\partial_0^{s-1}\right. \\
\left.+\la\lambda\,\hat{a}\ra\,a^{\alpha}\,\sum_{s=1}^{\infty}A_{\alpha\beta(2s-2)\dot\alpha}\,\lambda^{\beta(2s-2)}\,\partial_0^{s-1}\right]\hat{\sigma} + O\big(\la a\,\lambda\ra^{2s}\,\partial_0^{s-1}\big)\,,
\end{multline}
where ``$O(\la a\,\lambda\ra^{2s}\,\partial_0^{s-1})$'' denotes terms which obey the boundary conditions \eqref{bconds}. Removing the terms in \eqref{BSgauge4} which violate the boundary conditions imposes
\be\label{gfconstraints}
a^{\alpha}\,A_{\alpha\beta(2s-2)\dot\alpha}=0\,, \qquad \hat{a}^{\alpha}\,A_{\alpha\beta(2s-2)\dot\alpha}\,\lambda^{\beta(2s-2)}=\la a\,\lambda\ra^{2s-2}\,a^{\alpha}\,\partial_{\alpha\dot\alpha}\phi^{(s)}\,,
\ee
for each $s\geq1$. The solution to these constraints uniquely determines each HS-YM gauge potential:
\be\label{scalarpotential}
A_{\alpha(2s-1)\dot\alpha}(x)= a_{\alpha(2s-1)}\,a^{\beta}\,\partial_{\beta\dot\alpha}\phi^{(s)}(x)\,,
\ee
in terms of the adjoint-valued scalar function $\phi^{(s)}$ at each spin.

It is now possible to feed these expressions back into the holomorphic Chern-Simons action \eqref{hCS}, and integrate along the $\P^1$ fibres of twistor space to obtain an action functional on $\R^4$. The details of this computation are exactly the same as in the pure Yang-Mills case, so we refer the interested reader to~\cite{Bittleston:2020hfv}; following the steps in that calculation leads to the action:
\begin{multline}\label{ChalSieg}
S[\phi]=\frac{1}{2}\,\sum_{s=1}^{\infty}\,\int_{\R^{4}}\tr\left(\d\phi^{(s)}\wedge *\d\phi^{(s)}\right) \\
+\frac{1}{3}\sum_{s=1}^{\infty}\,\int_{\R^4}\mu_{a,a}\wedge\tr\left(\phi^{(s)}\,\sum_{r+t=s+1}\d\phi^{(r)}\wedge\d\phi^{(t)}\right)\,,
\end{multline}
where
\be\label{ASDform}
\mu_{a,a}:=a_{\alpha}\,a_{\beta}\,\d x^{\alpha\dot\alpha}\wedge\d x^{\beta}{}_{\dot\alpha}\,.
\ee
The equations of motion for each adjoint-valued scalar are
\be\label{scaleoms}
\Box\phi^{(s)}=-2\,a^{\alpha}\,a^{\beta}\,\sum_{r+t=s+1}\left[\partial_{\alpha}{}^{\dot\alpha}\phi^{(r)},\,\partial_{\beta\dot\alpha}\phi^{(t)}\right]\,,
\ee
which correspond precisely to the requirements that the HS-YM fields \eqref{scalarpotential} are self-dual.

This result can be summarized as the following:
\begin{thm}\label{SDThm}
The SD sector of HS-YM on $\R^4$ is equivalent an infinite tower of coupled, adjoint-valued scalars governed by the action \eqref{ChalSieg}. Furthermore, this theory is equivalent to a holomorphic Chern-Simons theory \eqref{hCS} on twistor space with volume form \eqref{poltf} and boundary conditions \eqref{bconds}, in the sense that extrema of the two actions are in one-to-one correspondence (up to gauge transformations).
\end{thm}

Observe that once again, the truncation to $s=1$ is self-consistent, in which case the gauge potential \eqref{scalarpotential} and action \eqref{ChalSieg} reduce to the Chalmers-Siegel description of SD Yang-Mills in terms of an adjoint-valued scalar field~\cite{Chalmers:1996rq}.

\medskip

We note that \eqref{scaleoms} are equivalent to the light-cone gauge description of SD HS-YM obtained in~\cite{Ponomarev:2017nrr}. To see this, one simply fixes $a^{\alpha}=(0,-1)$ and denotes\footnote{Here, $\partial$ and $\dbar$ are not to be confused with Dolbeault operators on twistor space. In the interest of matching the light-cone literature, we feel that this abuse of notation is momentarily acceptable.} 
\begin{align}
    \p^{0\dot0}=\p^+\,,\quad \p^{1\dot1}=\p^-\,,\quad \p^{0\dot1}=\bar\p\,,\quad \p^{1\dot0}=-\p\,,
\end{align}
so that $\Box=\p^+\p^-+\p\bar{\p}$. In this spin frame, \eqref{scaleoms} become
\begin{align}
    \Box \phi^{(s)}=2\,\sum_{r+t=s+1}\Big(\bar\p \phi^{(r)}\p^+\phi^{(t)}-\p^+\phi^{(r)}\bar{\p}\phi^{(t)}\Big)\,,
\end{align}
coinciding with the light-cont description of~\cite{Ponomarev:2017nrr}. We also note that the presence of the transverse derivative $\dbar$ on the right-hand side of this equation is a hallmark of locality~\cite{Metsaev:2005ar}.


\section{Twistor action for HS-YM}\label{sec:5}

Having established the classical integrability of the SD sector of HS-YM, we now turn to describing full HS-YM using twistor theory. This is possible because HS-YM admits a perturbative expansion around the SD sector, as evident when expressed in terms of a Lagrange multiplier field as in \eqref{ChSe1} -- \eqref{ChSe2}. It is fairly straightforward to construct the associated twistor action following the same recipe for pure Yang-Mills~\cite{Mason:2005zm,Boels:2006ir,Adamo:2011cb}. With the HS-YM twistor action in-hand, the tree-level MHV amplitudes are obtained by perturbatively expanding the portion of the action which encodes the non-SD interactions.


\subsection{Twistor action functional}

Let us now write down a twistorial description of full (classical) HS-YM by first recalling its spacetime action functional
\be\label{HS-YMact}
S[\bA,\mathbf{B}]=\sum_{s=1}^{\infty}\int_{\M}\d^{4}x\,\tr\left(B_{\alpha(2s)}\,F^{\alpha(2s)}\right) +\frac{\rg^2}{4}\,\int_{\M}\d^{4}x\,\tr\left(B_{\alpha(2s)}\,B^{\alpha(2s)}\right)\,.
\ee
The first set of terms in this action describes the SD sector, while the second set describes the linear non-SD fluctuations around SD HS-YM `background'. Using Theorem~\ref{HS-WC} and the Penrose transform, one infers that the first set corresponds to a holomorphic BF-action on twistor space:
\be\label{TA1}
S_{\mathrm{SD}}[\msf{a},\msf{b}]=\frac{\im}{2\pi}\int_{\PT}\D^{3}Z\wedge\tr\left(\msf{b}\wedge F^{(0,2)}[\msf{a}]\right)\,,
\ee
where $\msf{b}\in\Omega^{0,1}(\PT,\mathrm{End} E\otimes J^{\infty\,\vee}_{\P^1}\otimes\cO(-4))$. Expanded in $J^{\infty\,\vee}_{\P^1}$, the twistor field $\msf{b}$ is
\be\label{Tbfield}
\msf{b}=\sum_{s=1}^{\infty}b^{(s)}\,(e^0)^{s-1}\,, \qquad b^{(s)}\in\Omega^{0,1}(\PT,\mathrm{End}E\otimes\cO(-2s-2))\,,
\ee
where we recall that $e^0$ is `eaten' by inner product with $\p_0$ (i.e. $\p_0\intprod e^0=1$) in \eqref{TA1}, so that all projective scalings are respected. The resulting equations of motion for \eqref{TA1} on twistor space read
\be\label{TA1eq}
    F^{(0,2)}[\msf{a}]=0\,, \qquad \Dbar\msf{b}=0\,.
\ee
The first of these corresponds to the SD HS-YM equations, by virtue of Theorem~\ref{HS-WC}, while the second corresponds to the equations of motion of negative helicity HS-YM fields in a SD background. 

The latter follows from a non-abelian extension of the Penrose transform~\cite{Ward:1980am,Baston:1989vh}:
\be\label{naPenTran}
\left\{\bB_{\alpha\beta}\in\Omega^{2}_{-}\otimes\mathfrak{g}\otimes J^{\infty}_{\P^1} \mbox{ obeying } \bD^{\alpha\dot\alpha}\bB_{\alpha\beta}=0\right\} \cong H^{0,1}_{\Dbar}(\PT,\mathrm{End} V\otimes\cO(-4))\,,
\ee
where the HS-YM connection $\bD_{\alpha\dot\alpha}$ is assumed to be SD and $H^{0,1}_{\Dbar}$ is the Dolbeault cohomology group defined with respect to $\Dbar=\bar\partial+\msf{a}$ (which obeys $\Dbar^2=0$). Given a cohomology class in this group, the spacetime master field is constructed by an integral formula
\be\label{naPenTran1}
\bB_{\alpha\beta}(x|\lambda)=\int_{X}\D\lambda'\wedge\lambda'_{\alpha}\,\lambda'_{\beta}\,H^{-1}(x,\lambda')\,\msf{b}|_{X}\,H(x,\lambda')\,\sum_{s=1}^{\infty}\left(\la\lambda\,\lambda'\ra^2\,\partial_0\,e_0'\right)^{s-1}\,,
\ee
where\footnote{From now on, when $\la\lambda\,\d\lambda\ra$ serves as part of an integration measure, as opposed to a generator of $J^{\infty\,\vee}_{\P^1}$ which is always contracted away by inner products with generators of $J^{\infty}_{\P^1}$, it will always be denoted by $\D\lambda$ rather than $e^0$.} $\D\lambda'\equiv \la\lambda'\,\d\lambda'\ra$, $\lambda'$ is the homogeneous coordinate on $X\cong\P^1$ (which is integrated over) and $\lambda$ plays the role of the auxiliary parameter in the spacetime master field $\bB_{\alpha\beta}$. To see that this indeed solves the desired equation of motion, one uses the definition of the holomorphic frame, which implies $\lambda'_{\alpha}\bD^{\alpha\dot\alpha}H(x,\lambda')=0$, so that
\be\label{naPenTran2}
\begin{split}
\bD^{\alpha\dot\alpha}\bB_{\alpha\beta}(x|\lambda)&=\int_{X}\D\lambda'\wedge\lambda'_{\alpha}\,\lambda'_{\beta}\bD^{\alpha\dot\alpha}\,H^{-1}(x,\lambda')\,\msf{b}|_{X}\,H(x,\lambda')\,\sum_{s=1}^{\infty}\left(\la\lambda\,\lambda'\ra^2\,\partial_0\,e_0'\right)^{s-1} \\
&=\int_{X}\D\lambda'\wedge\lambda'_{\beta}\,H^{-1}(x,\lambda')\left(\lambda'_{\alpha}\partial^{\alpha\dot\alpha}\msf{b}|_{X}\right)H(x,\lambda')\,\sum_{s=1}^{\infty}\left(\la\lambda\,\lambda'\ra^2\,\partial_0\,e_0'\right)^{s-1}\\
&=0\,,
\end{split}
\ee
with the final equality following because $\lambda'_{\alpha}\partial^{\alpha\dot\alpha}\msf{b}|_{X}=0$ as a consequence of the incidence relations.

\medskip

The non-abelian twistor integral formula \eqref{naPenTran1} also suggests how to formulate the non-SD interactions of HS-YM non-locally on twistor space
\begin{multline}\label{TAint}
I[\msf{a},\msf{b}]=\int\limits_{\M\times\P^1\times\P^1}\!\!\d^{4}x\,\D\lambda_{1}\,\D\lambda_2\,\la\lambda_1\,\lambda_2\ra^2\,\cP_{12} \\
\times\,\tr\left[H^{-1}(x,\lambda_1)\,\msf{b}(x,\lambda_1)\,H(x,\lambda_1)\,H^{-1}(x,\lambda_2)\,\msf{b}(x,\lambda_2)\,H(x,\lambda_2)\right]\,,
\end{multline}
where the integral is taken over two copies of the same line in twistor space along with integration over the moduli space $\M$ of these lines. The object $\cP_{12}$ is a ``spin projector,'' valued in $J^{\infty}_{\P^1,1}\otimes J^{\infty}_{\P^{1},2}$ whose role is to absorb the factors of $J^{\infty\,\vee}_{\P^1}$ associated to each insertion of $\msf{b}$. It is defined by the requirements that it is holomorphic and has no scaling weight in $\lambda_1$ or $\lambda_2$.

The non-local twistor action \eqref{TAint} can be `compressed' further by denoting $\msf{b}_i\equiv\msf{b}(x,\lambda_i)$ and introducing the \emph{holomorphic Wilson line}~\cite{Mason:2010yk,Bullimore:2011ni}
\be\label{HWL}
U_{X}(\lambda_i,\lambda_j):=H(x,\lambda_i)\,H^{-1}(x,\lambda_j)\,,
\ee
associated to the partial connection $\Dbar=\dbar+\msf{a}$ on the bundle $V\to\PT$. These holomorphic Wilson lines act by parallel transport with respect to $\Dbar$ for which they are formal Green's functions on the twistor lines $X$:
\be\label{HWLcond}
U_{X}(\lambda_i,\lambda_j):\,V|_{X,\lambda_j}\rightarrow V|_{X,\lambda_i}\,, \qquad U_{X}(\lambda_i,\lambda_i)=\mathrm{id}_{\mathfrak{g}}\,,
\ee
where
\begin{equation*}
\Dbar|_{X_i}U_{X}(\lambda_i,\lambda_j)=\mathrm{id}_{\mathfrak{g}}\,\bar{\delta}(\la\lambda_i\,\lambda_j\ra)\,, \qquad \bar{\delta}(z):=\frac{1}{2\pi\im}\,\dbar\left(\frac{1}{z}\right)\,.
\end{equation*}
Here, $\mathrm{id}_{\mathfrak{g}}$ is the identity in the adjoint representation of the gauge group. 

In practical terms, the holomorphic Wilson loop can be represented as a path-ordered exponential
\be\label{HWL2}
U_{X}(\lambda_i,\lambda_j)=P\,\exp\left(-\int_{X}\omega_{ij}\wedge\msf{a}\right)\,,
\ee
where $\omega_{ij}$ is a meromorphic differential on $\P^1$ valued in $J^{\infty\,\vee}_{\P^1}$ with an infinite series of higher-order poles:
\be\label{omega}
\omega_{ij}(\lambda):=\frac{\D\lambda}{2\pi\im}\,\frac{\la\lambda_i\,\lambda_j\ra}{\la\lambda_i\,\lambda\ra\,\la\lambda\,\lambda_j\ra}\,\sum_{s=1}^{\infty}\left(\frac{e^0\,\la\lambda_i\,\lambda_j\ra}{\la\lambda_i\,\lambda\ra\,\la\lambda\,\lambda_j\ra}\right)^{s-1}\,.
\ee
The path-ordering symbol $P$ means that the holomorphic Wilson line can be expanded as an infinite series
\be\label{HWLseries}
U_{X}(\lambda_i,\lambda_j)=\mathrm{id}_{\mathfrak{g}}+\sum_{m=1}^{\infty}(-1)^m\,\prod_{k=1}^{m}\frac{\la\lambda_i\,\lambda_j\ra\,\D\lambda_k\,\msf{a}_k}{\la\lambda_{k-1}\,\lambda_k\ra\,\la\lambda_k\,\lambda_{k+1}\ra}\,\sum_{s_k=1}^{\infty}\left(\frac{e_k^0\,\la\lambda_i\,\lambda_j\ra}{\la\lambda_{i}\,\lambda_k\ra\,\la\lambda_k\,\lambda_{j}\ra}\right)^{s_k-1}\,,
\ee
where $\lambda_0\equiv\lambda_i$ and $\lambda_{m+1}\equiv\lambda_j$.

With these definitions, \eqref{TAint} becomes
\be\label{TAI}
I[\msf{a},\msf{b}]=\int\limits_{\M\times\P^1\times\P^1}\!\!\d^{4}x\,\D\lambda_{1}\,\D\lambda_2\,\la\lambda_1\,\lambda_2\ra^2\,\cP_{12}\,\tr\left[\msf{b}_1\,U_X(\lambda_1,\lambda_2)\,\msf{b}_2\,U_{X}(\lambda_2,\lambda_1)\right]\,,
\ee
and we define the full twistor action:
\be\label{TAct}
S[\msf{a},\msf{b}]=S_{\mathrm{SD}}[\msf{a},\msf{b}]+\frac{\rg^2}{4}\,I[\msf{a},\msf{b}]\,.
\ee
It is easy to see that this action is invariant under gauge transformations
\be\label{Tgtrans1}
\msf{a}\to\mathbf{g}\,\msf{a}\,\mathbf{g}^{-1}-\dbar\mathbf{g}\,\mathbf{g}^{-1}\,, \qquad \msf{b}\to\mathbf{g}\,\msf{b}\,\mathbf{g}^{-1}\,,
\ee
for any homogeneous function $\mathbf{g}(Z)$ on $\PT$ valued in $\mathfrak{g}\otimes J^{\infty}_{\P^1}$, since the holomorphic Wilson line transforms as
\be\label{HWLgt}
U_{X}(\lambda_1,\lambda_2)\to\mathbf{g}(x,\lambda_1)\,U_{X}(\lambda_1,\lambda_2)\,\mathbf{g}^{-1}(x,\lambda_2)\,.
\ee
The action also enjoys another local symmetry which acts only on the field $\msf{b}$:
\be\label{Tgtrans2}
\msf{b}\to\msf{b}+\Dbar \mathbf{f}\,, \qquad \mathbf{f}\in\Omega^0(\PT,\cO(-4)\otimes\mathfrak{g}\otimes J^{\infty\,\vee}_{\P^1})\,.
\ee
It is now possible to establish the following result:

\begin{thm}\label{TAThm}
The twistor action \eqref{TAct} is equivalent to HS-YM theory on $\R^4$, in the sense that solutions to its field equations are in one-to-one correspondence with solutions to the field equations of HS-YM (up to spacetime gauge transformations). Furthermore, the twistor action and HS-YM actions take the same value when evaluated on corresponding field configurations.
\end{thm}

\proof The proof follows exactly the same steps as in the construction of the twistor action for pure Yang-Mills theory~\cite{Mason:2005zm,Boels:2006ir}. The gauge freedom \eqref{Tgtrans1} -- \eqref{Tgtrans2} is used to put the twistor fields $\msf{a},\msf{b}$ into `harmonic' gauge
\be\label{hgauge}
\dbar^*|_{X}\msf{a}|_{X}=0=\dbar^{*}|_{X}\msf{b}|_{X}\,,
\ee
where $\dbar^*|_{X}$ is the adjoint of the $\dbar$-operator restricted to any twistor line. Since $\msf{a}$ and $\msf{b}$ are $(0,1)$-forms on $\PT$, it follows on dimensional grounds that $\dbar|_{X}\msf{a}|_{X}=0=\dbar|_{X}\msf{b}|_X$, so the gauge condition \eqref{hgauge} is equivalent to
\be\label{hgauge2}
\Delta_{X}\msf{a}|_X=0=\Delta_{X}\msf{b}|_X\,,
\ee
where $\Delta_X$ is the Laplacian on $\P^1$. As $H^{1}(\P^1,\cO)=0$, this implies that $\msf a|_{X}=0$. Residual gauge transformations must then respect 
\be\label{resg}
\begin{split}
 \dbar^*|_{X}\,\dbar|_{X}\mathbf{g}(Z)=0 & \qquad \Rightarrow \quad \mathbf{g}(Z)=\mathbf{g}(x|\lambda)\in\Omega^0(\R^4,\mathfrak{g}\otimes J^{\infty}_{\P^1})\,, \\
 \dbar^*|_{X}\,\dbar|_{X}\mathbf{f}(Z)=0 & \qquad \Rightarrow \quad \mathbf{f}(Z)=0\,,
\end{split}
\ee
with the last relation following from $H^0(\P^1,\cO(-4))=0$. This means that the residual gauge transformations are precisely the expected spacetime gauge transformations of HS-YM.

Now, in this gauge, the twistor fields can be expanded as~\cite{Woodhouse:1985id}
\begin{subequations}\label{WHrep}
    \begin{align}
        \msf{a}&=\msf{a}_{\dot\alpha}\,\bar{e}^{\dot\alpha}\,, \\
        \msf{b}&=\msf{b}_{\dot\alpha}\,\bar{e}^{\dot\alpha}+\sum_{s=1}^{\infty}(2s+1)\,\frac{B_{\alpha(2s)}(x)\,\hat{\lambda}^{\alpha(2s)}}{\la\lambda\,\hat{\lambda}\ra^{2s}}\,\bar{e}^0\,(e^0)^{s-1}\,,
    \end{align}
\end{subequations}
with the components $\msf{a}_{\dot\alpha}$, $\msf{b}_{\dot\alpha}$ as yet unconstrained. First, consider the portion of the action corresponding to $S_{\mathrm{SD}}$; evaluated on the fields \eqref{WHrep} in harmonic gauge this is:
\begin{multline}\label{Sdpart1}
S_{\mathrm{SD}}=\frac{\im}{2\pi}\int_{\PT}\frac{\D^{3}Z\wedge\D^{3}\hat{Z}}{\la\lambda\,\hat{\lambda}\ra^4}\,\tr\Bigg[\msf{b}_{\dot\alpha}\,\dbar_0\msf{a}^{\dot\alpha} \\
\left.+\sum_{s=1}^{\infty}(2s+1)\,\frac{B_{\alpha(2s)}\,\hat{\lambda}^{\alpha(2s)}}{\la\lambda\,\hat{\lambda}\ra^{2s}}\left(\dbar_{\dot\beta}a^{(s)\,\dot\beta}-\frac{1}{2}\sum_{r+t=s+1}[a^{(r)}_{\dot\beta},\,a^{(t)\,\dot\beta}]\right)\right]\,.
\end{multline}
Clearly, the field components $\msf{b}_{\dot\alpha}$ enter only as Lagrange multipliers. Integrating them out imposes $\dbar_0\msf{a}_{\dot\alpha}=0$, which by the usual extension of Liouville's theorem implies that
\be\label{LmultTA}
a^{(s)}_{\dot\alpha}=A_{\alpha(2s-1)\dot\alpha}(x)\,\lambda^{\alpha(2s-1)}\,, \qquad \mbox{for all }\,s\geq1\,.
\ee
The $\P^1$ degrees of freedom in \eqref{Sdpart1} can now be integrated out (cf., \cite{Boels:2006ir,Koster:2017fvf}), leaving the desired
\be\label{SDpart2}
S_{\mathrm{SD}}=\sum_{s=1}^{\infty}\int_{\R^4}\d^{4}x\,\tr\left(B_{\alpha(2s)}\,F^{\alpha(2s)}\right)\,,
\ee
for the SD part of the HS-YM action.

The non-local part of the twistor action is also easily evaluated in the harmonic gauge \eqref{WHrep}. Since $\msf{a}|_{X}=0$ in this gauge, the holomorphic Wilson lines become trivial, $U_{X}(\lambda_1,\lambda_2)=\mathrm{id}_{\mathfrak{g}}$, so the action reduces to 
\be\label{Intpart1}
I=\int\limits_{\R^{4}\times\P^1\times\P^1}\d^{4}x\,\D\lambda_1\,\D\lambda_2\,\la\lambda_1\,\lambda_2\ra^{2}\,\cP_{12}\,\tr\left(\msf{b}_1\,\msf{b}_{2}\right)\,.
\ee
Now, the requirements of homogeneity and holomorphicity uniquely fix the spin projector to be diagonal
\be\label{Dspinproj}
\cP_{12}=\sum_{s=1}^{\infty}\la\lambda_1\,\lambda_2\ra^{2s-2}\,\left(\partial_{0\,1}\,\partial_{0\,2}\right)^{s-1}\,,
\ee
which further simplifies the action to
\begin{multline}\label{Intpart2}
I=\sum_{s=1}^{\infty}\, (2s+1)^2\,\int\limits_{\R^4\times\P^1\times\P^1} \!\!\d^{4}x\,\frac{\D\lambda_1\wedge\D\hat{\lambda}_1}{\la\lambda_1\,\hat{\lambda}_1\ra^{2s+2}}\,\frac{\D\lambda_2\wedge\D\hat{\lambda}_2}{\la\lambda_2\,\hat{\lambda}_2\ra^{2s+2}}\,\la\lambda_1\,\lambda_2\ra^{2s} \\
\times\,\hat{\lambda}_{1}^{\alpha(2s)}\,\hat{\lambda}_{2}^{\beta(2s)}\,\tr\left(B_{\alpha(2s)}\,B_{\beta(2s)}\right)\,.
\end{multline}
Once again, the two $\P^1$ factors can be integrated out (cf., \cite{Boels:2006ir,Koster:2017fvf}), leaving
\be\label{Intpart3}
I=\sum_{s=1}^{\infty}\int_{\R^4}\d^{4}x\,\tr\left(B_{\alpha(2s)}\,B^{\alpha(2s)}\right)\,,
\ee
as desired.

This establishes that the twistor action \eqref{TAct} is literally equivalent to the spacetime HS-YM action in the harmonic gauge. Since the twistor action is itself gauge invariant, this completes the proof. \qed


\subsection{MHV amplitudes from twistor space}

It is natural to ask what a twistor description of HS-YM theory is actually good for. There are many potential answers to this question, but one that we pursue here is that twistor actions provide an easy way to obtain all-multiplicity scattering amplitude formulae for the MHV sector, as this is the first non-trivial scattering sector as we perturb away from self-duality. In particular, the classical generating functional for the tree-level MHV amplitudes is given by the non-local term in the twistor action, considered as a multi-linear functional of \emph{on-shell} (i.e., $\dbar$-closed) twistor fields -- see~\cite{Mason:2009afn,Adamo:2020yzi,Adamo:2022mev} for further explanation of this fact, which applies to any twistor action of the generic form $S_{\mathrm{SD}}+(\mbox{coupling})^2 I$. 

\medskip

Using the perturbative expansion of the holomorphic Wilson line \eqref{HWLseries}, this framework provides a twistorial formula for the $n$-point MHV amplitude:
\begin{multline}\label{TMHV1}
\cA_{n}^{\mathrm{MHV}}:=\cA_{n}(1^{+}_{s_1},\ldots,i^{-}_{s_i},\ldots,j^{-}_{s_j},\ldots,n^{+}_{s_n}) \\
=\int\d^{4}x\,\cP_{ij}|_{s_i,s_j}\,\frac{\la\lambda_i\,\lambda_j\ra^{6-n+\sum_{a\neq i,j}s_a}\,b^{(s_i)}_i\,b^{(s_j)}_j\,\D\lambda_i\,\D\lambda_j}{\la\lambda_1\,\lambda_2\ra\,\la\lambda_2\,\lambda_3\ra\cdots\la\lambda_n\,\lambda_1\ra} \prod_{b\neq i\,j}\frac{\D\lambda_b\,a^{(s_b)}_{b}}{\la\lambda_i\,\lambda_b\ra^{s_b-1}\,\la\lambda_b\,\lambda_j\ra^{s_b-1}}\,,
\end{multline}
where $\cP_{ij}|_{s_i,s_j}$ denotes the portion of the spin projector that selects the spins $s_i$ and $s_j$ for the negative helicity twistor representatives, ensuring that the integrand of this expression is homogeneous of degree zero in each point on the twistor line. 

Now, in the presence of additional insertions on $\P^1$, the spin projector is
\be\label{spinproj}
\cP_{ij}=\sum_{s_i,s_j=1}^{\infty}\partial_{0\,i}^{s_i-1}\,\partial_{0\,j}^{s_j-1}\,\tilde{\delta}(s_i-s_j)\,\tilde{\delta}\!\left(2-n+\sum_{a\neq i,j}s_a\right)\,\la\lambda_i\,\lambda_j\ra^{2s_i-2}\,,
\ee
as dictated by homogeneity, holomorphicity and gauge invariance in each of the positive helicity insertions. Feeding this into \eqref{TMHV1}, the spin constraints set all of the positive helicity external states to have spin-1, leaving
\be\label{TMHV2}
\cA_{n}^{\mathrm{MHV}}=\tilde{\delta}(s_i-s_j)\,\int\d^{4}x\,\frac{\la\lambda_i\,\lambda_j\ra^{2s_i+2}\,b^{(s_i)}_i\,b^{(s_j)}_j\,\D\lambda_i\,\D\lambda_j}{\la\lambda_1\,\lambda_2\ra\,\la\lambda_2\,\lambda_3\ra\cdots\la\lambda_n\,\lambda_1\ra} \prod_{b\neq i\,j}\D\lambda_b\,a^{(1)}_{b}\,,
\ee
as the expression fo the MHV amplitude on twistor space.

Now, to obtain a formula in momentum space we can simply evaluate \eqref{TMHV1} on momentum eigenstate representatives~\cite{Adamo:2011pv}:
\be\label{momeig}
\begin{split}
a^{(s_b)}_{b}&=\int_{\C^*}\frac{\d t_b}{t_b^{2s_b-1}}\,\bar{\delta}^{2}(\kappa_b-t_b\,\lambda)\,\e^{\im\,t_b\,[\mu\,b]}\,, \qquad b\neq i,j\,, \\
b^{(s_c)}_{c}&=\int_{\C^*}\d t_{c}\,t_{c}^{2s_c+1}\,\bar{\delta}^{2}(\kappa_c-t_c\,\lambda)\,\e^{\im\,t_c\,[\mu\,c]}\,, \qquad c= i,j\,,
\end{split}
\ee
where the holomorphic delta functions are defined by
\be\label{2dholdel}
\bar{\delta}^{2}(z):=\frac{1}{(2\pi\im)^2}\,\bigwedge_{\alpha=0,1}\dbar\left(\frac{1}{z_\alpha}\right)\,.
\ee
Inserting these representatives into \eqref{TMHV1} and using the explicit form of the spin projector \eqref{spinproj}, all of the integrals can be performed algebraically. In particular, the scale integrals in $t_b,t_c$ and $\P^1$ integrals are all performed against the holomorphic delta functions appearing in the twistor representatives, and the spacetime integration simply results in a momentum conserving delta function. This leaves
\be\label{TMHV3}
\cA_{n}^{\mathrm{MHV}}=(2\pi)^{4}\,\delta^{4}\!\left(\sum_{m=1}^{n}k_{m}\right)\,\tilde{\delta}(s_i-s_j)\,\frac{\la i\,j\ra^{2s_i+2}}{\la1\,2\ra\,\la2\,3\ra\cdots\la n\,1\ra}\,,
\ee
exactly matching the earlier claim \eqref{nMHV} for $n$-point MHV scattering in HS-YM.


\section{Discussion}\label{sec:6}

In this paper, we considered higher-spin Yang-Mills (HS-YM) theory: a non-abelian, chiral gauge theory with higher-spin degrees of freedom which extends previous constructions in the literature~\cite{Ponomarev:2017nrr,Krasnov:2021nsq,Tran:2021ukl} away from the purely self-dual sector. The theory has a complex action in real Lorentzian Minkowski spacetime, meaning that it is non-unitary and parity-violating, and its interaction vertices never contain more than a single spacetime derivative. Remarkably, these properties are enough for the theory to have non-vanishing higher-spin tree-level scattering amplitudes. The self-dual sector of the theory is classically integrable, and we used a twistor manifestation of this fact to explicitly construct the MHV tree-amplitudes of the theory.

While the non-unitarity and parity-violation of HS-YM are physically undesirable, it is surprising that such an otherwise fairly well-behaved theory (local, with only cubic and quartic interactions) has non-trivial scattering amplitudes. This contrasts with the widespread belief that non-trivial higher-spin scattering in flat spacetime requires some element of non-locality (cf., \cite{Ponomarev:2016lrm,Roiban:2017iqg,Taronna:2017wbx,Ponomarev:2017qab,Skvortsov:2018jea}); it seems that the ``get-out-of-jail-free'' card in the case of HS-YM is the intrinsic chirality of the fields, which enables an interacting theory to be constructed with only single-derivative vertices. Metaphorically speaking, if local higher-spin theories with trivial scattering amplitudes (such as chiral higher-spin gravity~\cite{Metsaev:1991mt,Metsaev:1991nb,Ponomarev:2016lrm} or self-dual HS-YM~\cite{Krasnov:2021nsq,Ponomarev:2017nrr}) live on a ``local island'' in the space of higher-spin theories, surrounded by a sea of non-local theories, then HS-YM lives on some sort of chiral \emph{buffer zone} between the two. This buffer zone is characterized by taking perturbative deformations of theories which live on the island; the HS-YM studied in this paper is clearly one example of a theory in the buffer zone, and there should be other examples, such as the higher-spin gauge theory induced by the IKKT matrix model~\cite{Ishibashi:1996xs,Steinacker:2022jjv}. It would be very interesting to generate further examples of such buffer zone theories, along with their explicitly non-vanishing scattering amplitudes.

\medskip

There are many other open questions and directions to explore following on from this work. In the first instance, another perspective on the restriction to spin-1 positive helicity degrees of freedom is desirable. Formulating HS-YM on the light-cone, where the need to make explicit choices for the polarization basis is removed, could shed light on this, as well as providing an independent check on our results.

All of the considerations in this paper have been classical; we have said nothing about the quantum consistency of HS-YM. A warm-up to answering this larger question would be to consider the quantum integrability of \emph{self-dual} HS-YM; it is expected that SD HS-YM will have non-vanishing all-positive helicity 1-loop scattering amplitudes that represent an anomaly to integrability, or equivalently, an anomaly in the twistor description of the self-dual sector~\cite{Bardeen:1995gk,Costello:2021bah,Bittleston:2022nfr}. In any case, this anomaly will boil down to a partition function-like calculation involving a sum over the degrees of freedom in the theory. Using zeta function regularization to treat the spectral sum (cf., \cite{Beccaria:2015vaa}), this will lead to $2\sum_{s=1}^{\infty}1=2\zeta(0)=-1$ and hence a non-vanishing anomaly. It would seem that an easy way to kill this anomaly would be to couple HS-YM with a complex scalar, with a term like $\bD^{\alpha\dot\alpha}\Phi\bD_{\alpha\dot\alpha}\bar{\Phi}$ in the Lagrangian. It would be interesting to consider this in more detail, both in spacetime and on twistor space.

It may also be interesting to explore HS-YM in the context of flat space, or celestial, holography (see~\cite{Pasterski:2021raf,McLoughlin:2022ljp} for reviews). For pure Yang-Mills theory and gravity, it has been shown that the classical infinite-dimensional symmetry algebras associated with the self-dual sectors have a natural manifestation on the celestial sphere in terms of local operators and their operator product expansions~\cite{Guevara:2021abz,Strominger:2021mtt}, and that these emerge naturally from the twistor descriptions of the self-dual sectors~\cite{Adamo:2021lrv}. Recently, it has been shown that Moyal deformations of the self-dual theories lead to enhancements of these classical symmetry algebras to their ``quantum'' deformations -- although the theories under consideration are still tree-level or 1-loop exact~\cite{Monteiro:2022lwm,Bu:2022iak,Guevara:2022qnm}. It has already been observed that these deformations are most naturally linked to a chiral higher-spin enhancement of the spacetime theories (rather than anything truly quantum mechanical), and self-dual HS-YM -- and its twistor description -- provides a first explicit realization of this fact. But more generally, it would be fascinating to explore how HS-YM and its scattering amplitudes fit into the recent proposals for higher-spin holography in asymptotically flat spacetimes~\cite{Ponomarev:2022ryp,Ponomarev:2022qkx}.

\acknowledgments
We thank Yannick Herfray, Kirill Krasnov, Zhenya Skvortsov and Harold Steinacker for helpful discussions, and Zhenya Skvortsov for comments on a draft. TT is grateful to the Asia Pacific Center for Theoretical Physics (APCTP) for hospitality during the mini workshop ``Higher Spin Gravity and its Applications'' during which the final version of this work was completed. TA is supported by a Royal Society University Research Fellowship and by the Leverhulme Trust (RPG-2020-386). The work of TT is partially supported by the Fonds de la Recherche Scientifique under Grants No. F.4503.20 (HighSpinSymm), Grant No. 40003607 (HigherSpinGraWave), T.0022.19 (Fundamental issues in extended gravitational theories) and the funding from the European Research Council (ERC) under Grant No. 101002551.

\appendix

\section{BCFW recursion}\label{app:A}

In this appendix, we prove the all-multiplicity formula for MHV scattering \eqref{nMHV} in HS-YM using on-shell recursion relations. While formally non-unitary, due to its complex Lagrangian, HS-YM is still a local, one-derivative field theory, which means that its tree-level scattering amplitudes will be meromorphic functions of the external kinematics with only simple poles corresponding to exchanged momenta. This means that tree-amplitudes of HS-YM could be recursively constructed using on-shell relations such as BCFW recursion~\cite{Britto:2005fq}. The only other required property (besides those already mentioned) is that the tree-amplitudes have sufficient fall-off under large values of the deformation parameter, $z$, used to define the recursion.

For the MHV helicity configuration of interest, it suffices to show that there is always a BCFW deformation under which the amplitude dies off at least as quickly as $z^{-1}$. Without loss of generality, we will consider $n$-point amplitudes for which particles $j$ and $n$ have negative helicity, with all others having positive helicity. Consider a deformation of the external kinematics for which
\be\label{BCFWdef}
\begin{split}
\kappa_1^{\alpha}\to\hat{\kappa}_{1}^{\alpha}(z)&:=\kappa_1^{\alpha}+z\,\kappa_{n}^{\alpha}\,, \\
\tilde{\kappa}_n^{\dot\alpha}\to\hat{\tilde{\kappa}}_{n}^{\dot\alpha}(z)&:=\tilde{\kappa}_{n}^{\dot\alpha}-z\,\tilde{\kappa}_{1}^{\dot\alpha}\,,
\end{split}
\ee
which maintains overall momentum conservation and keeps each external particle on-shell. Since particle $1$ has positive helicity and $n$ has negative helicity, we abbreviate this choice of deformation by calling it a ``$[-\,+\ra$-shift.'' A color-order partial scattering amplitude with deformed kinematics is denoted by $\cA_{n}(z)$. 

The structure of the deformation and the propagator \eqref{prop} of HS-YM theory ensures that, just as in ordinary Yang-Mills, the only $z$-dependence in any particular Feynman tree diagram contributing to $\cA_{n}(z)$ enters through the vertices and propagators which join the two deformed external legs. Any such propagator will have a denominator which depends on a subset of the external momenta, $I\subset\{1,\ldots,n\}$, containing one of $1$ or $n$ -- all other propagators are clearly independent of $z$. Without loss of generality, such a propagator will contribute a denominator of the form
\be\label{defdenom}
\hat{K}^{2}_{I}(z)=\left(\hat{k}_{1}(z)+\sum_{\substack{i\in I \\ i\neq j}}k_{i}\right)^2 = \left(\sum_{i\in I}k_i\right)^2-2\,z\,\la n|K_{I}|1]\,,
\ee
since momentum conservation can always be used to ensure that $1\in I$. Hence, the propagators of any diagram only contribute constant or $z^{-1}$ factors to the large-$z$ behavior of the amplitude. Interaction vertices can contribute, at most, factors of $z$, so the worst possible contribution to an amplitude, arising from $k$ cubic vertices linking the legs $1$ and $n$ with $k-1$ propagators, will scale linearly $z$ (this is the same argument as ordinary Yang-Mills~\cite{Britto:2005fq}).

The only other $z$-dependence comes from the external polarization vectors of particles $1$ and $n$. From \eqref{YMhel}, it is easy to see that these behave as
\begin{subequations}
\begin{align}\label{defpol}
\varepsilon^{(+)}_{1\,\alpha\dot\alpha}(z)&=\frac{\zeta_{1\,\alpha}\,\tilde{\kappa}_{1\,\dot\alpha}}{\la\hat{1}(z)\,\zeta\ra}\sim\frac{1}{z}\,, \\
\varepsilon^{(-)}_{n\,\alpha(2s_n-1)\dot\alpha}(z)&=\frac{\kappa_{n\,\alpha(2s_n-1)}\,\tilde{\zeta}_{n\,\dot\alpha}}{[\hat{n}(z)\,\tilde{\zeta}]}\sim\frac{1}{z}\,,
\end{align}
\end{subequations}
ensuring that $\cA_{n}(z)\sim z^{-1}$ as $|z|\to\infty$, as required. 

\medskip

We can now prove that the HS-YM MHV amplitude \eqref{nMHV} obeys the BCFW recursion relation using the $[-\,+\ra$-shift and induction. Just like in pure Yang-Mills, the MHV configuration is the homogeneous term in the recursion, meaning that only a single term, corresponding to pinching of a 3-point $\overline{\mathrm{MHV}}$ amplitude, contributes at each order in the recursion:
\begin{multline}\label{MHVrecur}
\cA_{n}^{\mathrm{MHV}}(1_{s_1}^+,2^+_{s_2},3^{+}_{1},\ldots,j^-_{s_j},\ldots,n^-_{s_n})= \cA_{3}^{\overline{\mathrm{MHV}}}\left(\hat{1}^+_{s_1}(z^*),2^+_{s_2},-\hat{K}^{-}_{1}(z^*)\right)\,\frac{1}{\la1\,2\ra\,[2\,1]} \\
\times\,\cA_{n-1}^{\mathrm{MHV}}\left(\hat{K}^+_{1}(z^*),3^+_{1},\ldots,j^-_{s_j},\ldots,\hat{n}^-_{s_n}(z^*)\right)\,,
\end{multline}
where
\be\label{ndeform1}
\hat{K}^{\alpha\dot\alpha}(z)=\kappa^{\alpha}_1\,\tilde{\kappa}^{\dot\alpha}_1+\kappa_2^{\alpha}\,\tilde{\kappa}^{\dot\alpha}_2+z\,\kappa_{n}^{\alpha}\,\tilde{\kappa}_1^{\dot\alpha}
\ee
and 
\be\label{zstar}
z^*:=-\frac{\la1\,2\ra}{\la n\,2\ra}\,,
\ee
is the critical value of the deformation parameter associated with this factorization channel. Note that we allow for generic external spins on the $\overline{\text{MHV}}$ factor, as the 3-point amplitude \eqref{MHV-bar} is well-defined even with higher-spin positive helicity legs.

We have already established that the formula \eqref{nMHV} is correct for $n=4$ by direct Feynman diagram calculations, so we proceed to use induction and assume that it also holds for $n-1$. With this inductive hypothesis, the constituents of \eqref{MHVrecur} become
\be\label{MHVbarrecur}
\cA_{3}^{\overline{\mathrm{MHV}}}\left(\hat{1}^+_{s_1}(z^*),2^+_{s_2},-\hat{K}^{-}_{1}(z^*)\right)=\tilde{\delta}(2-s_1-s_2)\,\frac{[1\,2]^{3}}{[\hat{K}\,1]^{2s_2-1}\,[2\,\hat{K}]^{2s_1-1}}\,,
\ee
and
\be\label{MHVrecur2}
\cA_{n-1}^{\mathrm{MHV}}\left(\hat{K}^+_{1}(z^*),3^+_{1},\ldots,j^-_{s_j},\ldots,\hat{n}^-_{s_n}(z^*)\right)=\tilde{\delta}(s_j-s_n)\,\frac{\la j\,n\ra^{2s_j+2}}{\la\hat{K}\,3\ra\cdots\la n-1\,n\ra\,\la n\,\hat{K}\ra}\,,
\ee
where $\hat{K}^{\alpha\dot\alpha}(z^*)\equiv\hat{K}^{\alpha}\,\hat{K}^{\dot\alpha}$ for
\be\label{deompmom}
\hat{K}^{\alpha}:=\frac{\la1\,n\ra}{\la2\,n\ra}\,\kappa_{2}^{\alpha}\,, \qquad \hat{K}^{\dot\alpha}:=\tilde{\kappa}_1^{\dot\alpha}+\frac{\la2\,n\ra}{\la1\,n\ra}\,\tilde{\kappa}_{2}^{\dot\alpha}\,.
\ee
Feeding all of this back into \eqref{MHVrecur} and observing that the spin constraint in \eqref{MHVbarrecur} sets $s_1=s_2=1$, we obtain 
\be\label{MHVrecur3}
\cA_{n}^{\mathrm{MHV}}=\tilde{\delta}(s_j-s_n)\,\frac{\la j\,n\ra^{2s_j+2}}{\la1\,2\ra\,\la2\,3\ra\cdots\la n\,1\ra}\,,
\ee
which is precisely the desired formula.

\setstretch{0.7}
\footnotesize
\bibliography{twistor}
\bibliographystyle{JHEP}

\end{document}